\documentclass[%
reprint,
superscriptaddress,
amsmath,amssymb,
aps,
]{revtex4-2}
\usepackage{array} % Required for custom column specifications
\usepackage{booktabs} % Provides better-looking horizontal lines
\usepackage{comment}
\usepackage{xcolor}
\usepackage{graphicx}% Include figure files
\usepackage{dcolumn}% Align table columns on decimal point
\usepackage{bm}% bold math
\usepackage{hyperref}% add hypertext capabilities
\begin{document}

\preprint{APS/123-QED}

%\begin{frontmatter}

%\title{Identifiying the Limit of Proton Binding for Tm Isotopes \tnoteref{mytitlenote}}
\title{Convergence on the Proton Drip-Line in Thulium
%\title{Boundary of the Proton Drip-Line for Thulium Isotopes
%\title{The Proton Drip-Line Boundary for Thulium Isotopes
%\title{Finding the Proton Drip-Line Boundary in Thulium
%"Defining the Threshold of the Proton Drip-Line in Thulium"
%"Finding the Proton Drip-Line Boundary in Thulium"
%\title{Establishing the Threshold of the Proton Drip-Line in Thulium
%\title{Determining The Last Proton-Bound Thulium isotope
%\title{Staking out the Proton Drip-Line of Thulium
%\\Tm and Er masses cement the $Z=69$ proton drip-line\\Pinning Down the Proton Drip-Line of Tm
} 

\author{B. Kootte}%[$^{\textrm{\textdagger}}$]
\thanks{Corresponding author (brian.a.kootte@jyu.fi)}
%\affiliation{Department of Physics and Astronomy, University of Manitoba, Winnipeg, MB, R3T 2N2, Canada}
\affiliation{Department of Physics and Astronomy, University of Manitoba, 30A Sifton Road, Winnipeg, MB, R3T 2N2, Canada}
\affiliation{TRIUMF, 4004 Wesbrook Mall, Vancouver, BC, V6T 2A3, Canada}

\author{M.P. Reiter}%7?
\affiliation{TRIUMF, 4004 Wesbrook Mall, Vancouver, BC, V6T 2A3, Canada}
%\affiliation{II. Physikalisches Institut, Justus-Liebig-Universität, 35392 Gießen, Germany}
\affiliation{II. Physikalisches Institut, Justus-Liebig-Universität, Heinrich-Buff-Ring 16, 35392 Gießen, Germany}
%\affiliation{School of Physics and Astronomy, University of Edinburgh, Edinburgh EH9 3FD, Scotland, UK}
\affiliation{School of Physics and Astronomy, University of Edinburgh, James Clerk Maxwell Building, Peter Guthrie Tait Road, Edinburgh EH9 3FD, Scotland, UK}

%Yb Beamtime Author:
\author{C. Andreoiu}
%\affiliation{Department of Chemistry, Simon Fraser University, Burnaby, BC, V5A 1S6, Canada}
\affiliation{Department of Chemistry, Simon Fraser University, 8888 University Drive, Burnaby, BC, V5A 1S6, Canada}

%Yb Beamtime Author:
\author{S. Beck}
%\affiliation{II. Physikalisches Institut, Justus-Liebig-Universität, 35392 Gießen, Germany}
\affiliation{II. Physikalisches Institut, Justus-Liebig-Universität, Heinrich-Buff-Ring 16, 35392 Gießen, Germany}
%\affiliation{GSI Helmholtzzentrum für Schwerionenforschung GmbH, 64291 Darmstadt, Germany}
\affiliation{GSI Helmholtzzentrum für Schwerionenforschung GmbH, Planckstraße 1, 64291 Darmstadt, Germany}

%Yb Beamtime Author:
\author{J. Bergmann}
%wrong!%\affiliation{Université de Strasbourg, CNRS, IPHC UMR 7178, F-67 000 Strasbourg, France}
%?
\affiliation{II. Physikalisches Institut, Justus-Liebig-Universität, Heinrich-Buff-Ring 16, 35392 Gießen, Germany}
%?

%Yb Beamtime Author:
\author{T. Brunner} 
%\affiliation{Physics Department, McGill University, Montréal, QC, H3A 2T8, Canada}
\affiliation{Physics Department, McGill University, 3600 rue University, Montréal, QC, H3A 2T8, Canada}

%Yb Beamtime Author:
%%\author{Dominique Curien$^{23}$} 
%Yb Beamtime Author:
%%\author{Irene Dedes$^{16,17}$}
%Yb Beamtime Author:
\author{T. Dickel}
%\affiliation{II. Physikalisches Institut, Justus-Liebig-Universität, 35392 Gießen, Germany}
\affiliation{II. Physikalisches Institut, Justus-Liebig-Universität, Heinrich-Buff-Ring 16, 35392 Gießen, Germany}
%\affiliation{GSI Helmholtzzentrum für Schwerionenforschung GmbH, 64291 Darmstadt, Germany}
\affiliation{GSI Helmholtzzentrum für Schwerionenforschung GmbH, Planckstraße 1, 64291 Darmstadt, Germany}

\author{K.A. Dietrich}%
%\affiliation{Ruprecht-Karls-Universität Heidelberg, D-69117 Heidelberg, Germany}
%\affiliation{Ruprecht-Karls-Universität Heidelberg, Grabengasse 1, 69117 Heidelberg, Germany}
%or
%Fakultät für Physik und Astronomie
%Im Neuenheimer Feld 226
%69120 Heidelberg
%Or
\affiliation{Fakultät für Physik und Astronomie,
Ruprecht-Karls-Universität Heidelberg,
Im Neuenheimer Feld 226,
69120 Heidelberg, Germany}
\affiliation{TRIUMF, 4004 Wesbrook Mall, Vancouver, BC, V6T 2A3, Canada}

%Yb Beamtime Author:
\author{J. Dilling}%\author{Jens Dilling$^{14,19,??}$}
\affiliation{Oak Ridge National Laboratory
1 Bethel Valley Road, Oak Ridge, TN, 37830, USA}
\affiliation{Physics Department, Duke University, 120 Science Drive, 
Campus Box 90305 
Durham, NC, 27708, 
USA}

%Yb Beamtime Author:
%%\author{Jerzy Dudek$^{23,17}$}
%Yb Beamtime Author?:
%\author{I: Dillmann$^{2}$}
%
\author{E. Dunling}
\affiliation{TRIUMF, 4004 Wesbrook Mall, Vancouver, BC, V6T 2A3, Canada}

%Yb Beamtime Author:
\author{J. Flowerdew}
%\affiliation{Department of Physics and Astronomy, University of Calgary, Calgary, AB, T2N 1N4, Canada}
\affiliation{Department of Physics and Astronomy, University of Calgary, 2500 University Drive NW, Calgary, AB, T2N 1N4, Canada}

%Yb Beamtime Author:
%%\author{Abdelghafar Gaamouci$^{26}$} 
%Yb Beamtime Author:
\author{L. Graham}
\affiliation{TRIUMF, 4004 Wesbrook Mall, Vancouver, BC, V6T 2A3, Canada}

\author{G. Gwinner}
%\affiliation{Department of Physics and Astronomy, University of Manitoba, Winnipeg, MB, R3T 2N2, Canada}
\affiliation{Department of Physics and Astronomy, University of Manitoba, 30A Sifton Road, Winnipeg, MB, R3T 2N2, Canada}

\author{Z. Hockenbery}
%\affiliation{Physics Department, McGill University, Montréal, QC, H3A 2T8, Canada}
\affiliation{Physics Department, McGill University, 3600 rue University, Montréal, QC, H3A 2T8, Canada}
\affiliation{TRIUMF, 4004 Wesbrook Mall, Vancouver, BC, V6T 2A3, Canada}

\author{C. Izzo}%
\affiliation{TRIUMF, 4004 Wesbrook Mall, Vancouver, BC, V6T 2A3, Canada}

\author{A. Jacobs} %\author{A. Jacobs$^{2}$}
\affiliation{Department of Physics and Astronomy, University of British Columbia, 6224 Agricultural Road, Vancouver, BC, V6T 1Z1, Canada}
\affiliation{TRIUMF, 4004 Wesbrook Mall, Vancouver, BC, V6T 2A3, Canada}

\author{A. Javaji}
\affiliation{Department of Physics and Astronomy, University of British Columbia, 6224 Agricultural Road, Vancouver, BC, V6T 1Z1, Canada}
\affiliation{TRIUMF, 4004 Wesbrook Mall, Vancouver, BC, V6T 2A3, Canada}

%Yb Beamtime Author:
\author{R. Klawitter}
%\affiliation{Ruprecht-Karls-Universität Heidelberg, D-69117 Heidelberg, Germany}
%\affiliation{Ruprecht-Karls-Universität Heidelberg, Grabengasse 1, 69117 Heidelberg, Germany}
\affiliation{Fakultät für Physik und Astronomie,
Ruprecht-Karls-Universität Heidelberg,
Im Neuenheimer Feld 226,
69120 Heidelberg, Germany}
\affiliation{TRIUMF, 4004 Wesbrook Mall, Vancouver, BC, V6T 2A3, Canada}

%Yb Beamtime Author:
\author{Y. Lan}
\affiliation{Department of Physics and Astronomy, University of British Columbia, 6224 Agricultural Road, Vancouver, BC, V6T 1Z1, Canada}
\affiliation{TRIUMF, 4004 Wesbrook Mall, Vancouver, BC, V6T 2A3, Canada}

\author{E. Leistenschneider}
\affiliation{Department of Physics and Astronomy, University of British Columbia, 6224 Agricultural Road, Vancouver, BC, V6T 1Z1, Canada}
\affiliation{TRIUMF, 4004 Wesbrook Mall, Vancouver, BC, V6T 2A3, Canada}

\author{E.M. Lykiardopoulou}
\affiliation{Department of Physics and Astronomy, University of British Columbia, 6224 Agricultural Road, Vancouver, BC, V6T 1Z1, Canada}
\affiliation{TRIUMF, 4004 Wesbrook Mall, Vancouver, BC, V6T 2A3, Canada}

%Yb Beamtime author
\author{I. Miskun}%
%Ivan Miskun?% Physikalisches Institut, Justus-Liebig-Universität Gießen, 35392 Gießen, Germany
%\affiliation{II. Physikalisches Institut, Justus-Liebig-Universität, 35392 Gießen, Germany}
\affiliation{II. Physikalisches Institut, Justus-Liebig-Universität, Heinrich-Buff-Ring 16, 35392 Gießen, Germany}

\author{I. Mukul}%
\affiliation{TRIUMF, 4004 Wesbrook Mall, Vancouver, BC, V6T 2A3, Canada}

\author{T. Murb\"{o}ck}%
\affiliation{TRIUMF, 4004 Wesbrook Mall, Vancouver, BC, V6T 2A3, Canada}

%Yb Beamtime Author:
%%\author{Nikolay Minkov,27} 
%Yb Beamtime Author:
%\author{Victor Monier$^{1}$}
%

\author{S.F. Paul}
%\affiliation{Ruprecht-Karls-Universität Heidelberg, D-69117 Heidelberg, Germany}
%\affiliation{Ruprecht-Karls-Universität Heidelberg, Grabengasse 1, 69117 Heidelberg, Germany}
\affiliation{Fakultät für Physik und Astronomie,
Ruprecht-Karls-Universität Heidelberg,
Im Neuenheimer Feld 226,
69120 Heidelberg, Germany}
\affiliation{TRIUMF, 4004 Wesbrook Mall, Vancouver, BC, V6T 2A3, Canada}

%Yb Beamtime Author:
\author{W.R. Plaß}
%\affiliation{II. Physikalisches Institut, Justus-Liebig-Universität, 35392 Gießen, Germany}
\affiliation{II. Physikalisches Institut, Justus-Liebig-Universität, Heinrich-Buff-Ring 16, 35392 Gießen, Germany}
%\affiliation{GSI Helmholtzzentrum für Schwerionenforschung GmbH, 64291 Darmstadt, Germany}
\affiliation{GSI Helmholtzzentrum für Schwerionenforschung GmbH, Planckstraße 1, 64291 Darmstadt, Germany}
%\author{M.P. Reiter$^{2,7,12}$}%7?

\author{J. Ringuette}%
\affiliation{Department of Physics, Colorado School of Mines, 1500 Illinois St., Golden, CO, 80401, USA}
\affiliation{TRIUMF, 4004 Wesbrook Mall, Vancouver, BC, V6T 2A3, Canada}

%Yb Beamtime Author:
\author{C. Scheidenberger}
%\affiliation{II. Physikalisches Institut, Justus-Liebig-Universität, 35392 Gießen, Germany}
\affiliation{II. Physikalisches Institut, Justus-Liebig-Universität, Heinrich-Buff-Ring 16, 35392 Gießen, Germany}
%\affiliation{GSI Helmholtzzentrum für Schwerionenforschung GmbH, 64291 Darmstadt, Germany}
\affiliation{GSI Helmholtzzentrum für Schwerionenforschung GmbH, Planckstraße 1, 64291 Darmstadt, Germany}
%\affiliation{Helmholtz Forschungsakademie Hessen für FAIR (HFHF), GSI Helmholtzzentrum für Schwerionenforschung, Campus Gießen, 35392 Gießen, Germany}
\affiliation{Helmholtz Forschungsakademie Hessen für FAIR (HFHF), GSI Helmholtzzentrum für Schwerionenforschung, Campus Gießen, Heinrich-Buff-Ring 16, 35392 Gießen, Germany}

\author{R. Silwal}
\affiliation{TRIUMF, 4004 Wesbrook Mall, Vancouver, BC, V6T 2A3, Canada}
%\affiliation{Department of Physics and Astronomy, Appalachian State University, Boone, NC, 28607, USA}
\affiliation{Department of Physics and Astronomy, Appalachian State University, 231 Garwood Hall, 525 Rivers Street, Boone, NC, 28608, USA}

\author{R. Simpson}
\affiliation{TRIUMF, 4004 Wesbrook Mall, Vancouver, BC, V6T 2A3, Canada}
\affiliation{Department of Physics and Astronomy, University of British Columbia, 6224 Agricultural Road, Vancouver, BC, V6T 1Z1, Canada}

%Yb Beamtime Author:
\author{A. Teigelh\"ofer}%Andrea Teigelh\"{o}fer
\affiliation{TRIUMF, 4004 Wesbrook Mall, Vancouver, BC, V6T 2A3, Canada}

\author{R.I. Thompson}
%\affiliation{Department of Physics and Astronomy, University of Calgary, Calgary, AB, T2N 1N4, Canada}
\affiliation{Department of Physics and Astronomy, University of Calgary, 2500 University Drive NW, Calgary, AB, T2N 1N4, Canada}

%Yb Beamtime Author:
\author{J.L. Tracy, Jr.} 
\affiliation{TRIUMF, 4004 Wesbrook Mall, Vancouver, BC, V6T 2A3, Canada}
%Yb Beamtime Author:
\author{M. Vansteenkiste} %{Michael Vansteenkiste 
\affiliation{TRIUMF, 4004 Wesbrook Mall, Vancouver, BC, V6T 2A3, Canada}

%Yb Beamtime Author:
%\author{Hua-Lei Wang$^{29}$} 
%
\author{R. Weil}%%\author{R. Weil$^{3,8}$}%Quantum matter institute
\affiliation{TRIUMF, 4004 Wesbrook Mall, Vancouver, BC, V6T 2A3, Canada}
\affiliation{Department of Physics and Astronomy, University of British Columbia, 6224 Agricultural Road, Vancouver, BC, V6T 1Z1, Canada}

%Yb Beamtime Author:
\author{M.E. Wieser}%Michael E. Wieser
%\affiliation{Department of Physics and Astronomy, University of Calgary, Calgary, AB, T2N 1N4, Canada}
\affiliation{Department of Physics and Astronomy, University of Calgary, 2500 University Drive NW, Calgary, AB, T2N 1N4, Canada}

%Yb Beamtime Author:
\author{C. Will}%Christian Will 
%\affiliation{II. Physikalisches Institut, Justus-Liebig-Universität, 35392 Gießen, Germany}
\affiliation{II. Physikalisches Institut, Justus-Liebig-Universität, Heinrich-Buff-Ring 16, 35392 Gießen, Germany}

%Yb Beamtime Author:
%\author{Jie Yang$^{17,29}$}

\author{A.A. Kwiatkowski}%\\
\affiliation{TRIUMF, 4004 Wesbrook Mall, Vancouver, BC, V6T 2A3, Canada}
\affiliation{Department of Physics and Astronomy, University of Victoria, 3800 Finnerty Road, Victoria, BC, V8P 5C2, Canada}

\date{\today}% It is always \today, today,

\begin{abstract}

%Measurements were performed using the Multiple Reflection, Time-Of-Flight Mass Spectrometer (MR-TOF-MS) at TRIUMF's TITAN facility, and were made possible by utilizing a technique recently adapted from stable isotope measurements to mass spectra of radioactive isotopes known as \emph{Mass selective retrapping} \cite{REITER2021MRTOF_commissioning, dickel2017isobar}.

%, and were made possible by utilizing a technique known as \emph{Mass selective retrapping} \cite{REITER2021MRTOF_commissioning, dickel2017isobar} which has been adapted from stable isotope measurements to mass spectra of radioactive isotopes.
%This technique allowed us to access the relevant isotopes by suppressing the far more abundant species which dominates the radioactive beam at these masses, and to help to maintain favourable isotope ratios.
%%maintain favourable ratios of isotopes.

%Measurements were performed using the Multiple Reflection, Time-Of-Flight Mass Spectrometer (MR-TOF-MS) at TRIUMF's TITAN facility to explore the limit of proton-bound Tm.

Direct observation of proton emission for very small $Q$·-values is often unfeasible due to the long partial half-lives of the proton emission channel associated with tunneling through the Coulomb barrier.
Therefore, proton emitters with very small decay energies may require the masses of both parent and daughter nuclei in order to establish them as proton unbound.
%For the weakest of proton emitters we therefore rely on nuclear masses to determine the Q-value of a candidate proton emitter.
%
Nuclear mass models have been used to predict the proton drip-line of the thulium (Tm) isotopic chain ($Z=69$), but until now the proton separation energy has not been experimentally tested. 
%The proton drip-line of the Tm isotopic chain ($Z=69$) remains experimentally undetermined, with only models predicting the last proton-bound isotope.
%
Mass measurements were performed using a Multiple Reflection Time-Of-Flight Mass Spectrometer (MR-TOF-MS) at TRIUMF's TITAN facility to conclusively map the limit of proton-bound Tm.
The masses of neutron-deficient, $^{149}$Tm and $^{150}$Tm, combined with measurements of $^{149m,g}$Er (which were found to deviate from literature by $\approx$150 keV), provide the first experimental confirmation that $^{149}$Tm is the first proton-unbound nuclide in the Tm chain. Our measurements also enable the strength of the $N=82$ neutron shell gap to be determined at the Tm proton drip-line, providing evidence supporting its continued existence. 
%where an abrupt change in the deformation is known to occur \cite{delion2021universal}.

%These measurements unambiguously identify the proton drip-line to lie between proton-bound $^{150}$Tm and proton-unbound $^{149}$Tm. 
%The experimental confirmation of $^{149}$Tm as a proton emitter also allows a prediction to be made of the proton emission partial half-life.

\end{abstract}

%\maketitle

%\begin{keyword}
%\texttt{elsarticle.cls}\sep proton emission \sep proton drip-line \sep template
%\MSC[2010] 00-01\sep  99-00
%\end{keyword}

%\linenumbers

%\twocolumn

\maketitle

%\clearpage
\section{Introduction}

%As isotopes become increasingly neutron-deficient they lose the stabilizing attractive nuclear force provided by the neutrons until it becomes energetically possible for one or more protons to spontaneously escape the nucleus.
When isotopes become sufficiently neutron-deficient it eventually becomes energetically possible for one or more protons to spontaneously escape the nucleus.
This transition within an isotopic chain to one or more protons being unbound is known as the \emph{proton drip-line}, and its location is of fundamental interest.
It can be experimentally determined through measurements of the one-proton separation energy, $S_{p}$, 
%which is defined as $S(p) = -M(A,Z) + M(A-1,Z-1) + m_p$
and a nucleus is said to lie beyond the proton drip-line when $S_{p} \leq 0$ (i.e. it is \emph{proton-unbound}) \cite{WoodsP.J.1997NBTP,PfutznerM.2012Rdal}.
Studying the location of the proton drip-line provides a valuable benchmark for the various nuclear models that predict the properties of unstable nuclei. 

%and single-proton spectroscopy can provide a useful probe of single-particle structure.

%[One of the most compelling frontiers of nuclear physics lies in the study of systems far from stability, the study of which requires state-of-the-art production mechanisms to be combined with high-sensitivity measurement techniques.]
%An interesting transition in the decay properties of unstable ground state nuclei is the crossing of the drip-lines. 

%^1_1$H
%The proton separation energy is given by $S(p) = -M(A,Z) + M(A-1,Z-1) + ^1_1$H and when it becomes energetically favourable for a single proton to escape a nucleus ($S_{p} \leq 0$) that species is said to lie beyond the proton drip-line or to be \emph{proton-unbound} \cite{WoodsP.J.1997NBTP}\cite{PfutznerM.2012Rdal}.

%it is possible for species at or beyond the proton drip-line to have significantly longer half-lives than their counterparts at the neutron drip-line due to the repulsive Coulomb potential just outside the nucleus.
%In some species the energy of the decay ($Q_p$) is so low that...
%classically forbidden
Near the proton drip-line, a proton must tunnel through the Coulomb barrier in order to be emitted (i.e. a single-nucleon analogue to $\alpha$-decay).
%In order for a proton to be emitted, it first must tunnel through the Coulomb barrier (i.e. a single-nucleon analogue to $\alpha$-decay). 
%This quantum tunneling process is analogous to $\alpha$-decay, but doesn't require the formation of a helium nucleus to tunnel through the potential barrier.
%However the Coulomb barrier can extend the ground state proton-emission lifetimes of proton-unbound nuclides well beyond the time scale of competing decay modes (order of seconds for many drip-line nuclei), making proton emission near the drip-line extremely challenging to observe directly.
%The long half-lives of many proton emitting nuclei resulting from the Coulomb barrier suppression of this decay mode allows very weakly-bound or even unbound species to be studied using high-precision techniques.
%Compared to alpha decay, calculations of the proton emission decay rate or spectroscopic factor are simplified for a single emitted nucleon since it doesn't first require the formation of a cluster of nucleons \cite{WoodsP.J.1997NBTP}.
In contrast to $\alpha$ decay, however, it is simpler to calculate the proton emission decay rate or spectroscopic factor, since the decay does not require the formation of a cluster of nucleons \cite{WoodsP.J.1997NBTP}. 
%Compared to alpha decay, calculating the proton emission decay rate or spectroscopic factor is simpler for a single emitted nucleon, as it doesn't require forming a cluster of nucleons \cite{WoodsP.J.1997NBTP}. 
%
Proton emission is significantly influenced by the centrifugal barrier, and
the proton emission half-life at a given proton number $Z$ depends strongly 
%more sensitively 
on underlying nuclear structure, including the angular momentum carried away by the emitted proton \cite{WoodsP.J.1997NBTP}\cite{Delion2006PRL}\cite{Zhang2018} and the decay energy $Q_p$ \cite{Zhang2018}.

Furthermore, the nuclear shell structure can change towards the drip-line; effects such as the weakening or disappearance of shells (quenching) at the classical magic numbers and the appearance of new magic numbers have been both predicted and observed \cite{SORLIN2008602, Kanungo_2013}. 
Nuclear masses and binding energies provide indispensable information when studying the limits of nuclear existence and shell structure.
They allow calculations of the particle separation energies and can be used to make predictions regarding the underlying nuclear structure and nuclear properties such as half-life, angular momentum, or spectroscopic factors in the case of proton emission.

The mass values of nuclides across the $N=82$ shell closure, formed between the $h_{11/2}$ (or nearby $s_{1/2}$ or $d_{3/2}$) and the $f_{7/2}$ orbitals, have been intensively studied on the neutron-rich side of the nuclear chart; see, e.g., \cite{dworschak2008restoration_N82,breitenfeldt2010approach_N82,Atanasov2015_N82,knoebel2016_N82,Babcock2018,PhysRevLett.124.092502}. 
Corresponding investigations at the neutron-deficient $N=82$ shell closure up to the elements of holmium ($Z=67$) and erbium ($Z=68$) have been performed at ISOLTRAP (ISOLDE/CERN) using Penning trap \cite{beck2000accurate}, at the ESR storage ring (GSI) using Schottky mass spectrometry \cite{2005Li24}, and recently for ytterbium ($Z=70$) at TITAN \cite{BeckSoenke2021_Yb} using the multiple reflection time-
of-flight mass spectrometer (MR-TOF-MS) technique \cite{WOLLNIK_MRTOF_1990}, which is well suited for measurements of short-lived isotopes.
The proton drip-line around $Z=69$ is associated with an abrupt change in the deformation \cite{delion2021universal, WoodsP.J.1997NBTP, geng2004proton}. 
%[cite Woods et al., ARNPS 47 (1997)541, Geng et al., Prog Theo Phys 112 (2004) 4]
Although masses of neutron-deficient thulium isotopes have been measured by SHIPTRAP \cite{Rauth2008_Penning_beyond_p_drip}, measurements remain elusive across the $N=82$ shell closure where the one-proton drip-line is predicted to lie.  
Using the MR-TOF-MS at TITAN, here we report the first high-precision mass measurements of neutron-deficient thulium isotopes at the proton drip-line, enabling S$_{2N}$ to be determined across $N=82$.

\subsection{Background}

Ground state proton emission can occur for any nucleus beyond the proton drip-line, and was first observed in 1982 in $^{151}$Lu and $^{147}$Tm \cite{hofmann1982proton, klepper1982direct}.
Since then, at least 27 heavy ground state proton emitters have been identified, along with a number of isomeric states \cite{BLANK2008403}\cite{ENSDF-webpage}.

While $^{147}$Tm has long been known to be a proton emitter, and prior direct mass measurements of $^{147}$Tm and $^{148}$Tm have been able to constrain the location of the proton drip-line to lie somewhere between $A=148$ and $A=151$ \cite{Rauth2008_Penning_beyond_p_drip}, the precise isotope at which Tm becomes proton-unbound remains uncertain due to the thus far unmeasured masses of $^{149}$Tm and $^{150}$Tm.
In cases where neither observation of the decay nor direct mass measurements can be used to establish the drip-line, various theoretical approaches are employed to predict its location, but these models are known to struggle to predict $S_p$ far from stability \cite{zhang2022observation}.
%phenomenological descriptions such as LDM...?

Density functional theory and phenomenological descriptions of the nuclear binding energy have been utilized to predict the location of the proton drip-line.
For instance, the Skyrme results of \cite{ERLERJochen2012Tlot} %\footnotemark \footnotetext{test}
%\footnote{test}
%(available at \href{http://massexplorer.frib.msu.edu}{massexplorer.frib.msu.edu}) 
can be used to interpolate the $S_p$ of odd-Z nuclei. 
However, for $^{149}$Tm, the choice of energy density functional -- such as UNEDF1 \cite{Kortelainen2012UNEDF1} or SV-min \cite{Klupfel2009_SV-min} -- leads to competing predictions regarding its position relative to the drip-line. %whether or not it is a drip line nucleus.
Furthermore, the newer BSkG3 model \cite{BSkG3_grams2023skyrme} predicts the drip-line to be even closer to stability.

Bayesian model averaging can be used to assign weights to models in an attempt to improve estimates of the binding energies of heavy nuclei \cite{Kejzlar2020statisticalBE}.
A recent Bayesian statistical analysis was performed to weight the proton drip-line predictions provided by several different energy density functionals, as well as FRDM2012 and HFB-24  \cite{Tm_drip_2020}.
This work predicted the last proton-bound isotope of Tm to occur at $N=81$ ($A=150$), but this has never been experimentally verified.% through either a measurement of the proton separation energy ($S_p$), or otherwise.
The precise location of the thulium proton drip-line remains a significant gap in our experimental knowledge of the heavy, proton-deficient nuclei.
The present measurement aims to determine the exact position of the ground state proton drip-line for Tm ($Z=69$).

\section{Experimental Description} 

%Data was collected over two separate experiments which are treated separately here due to large differences in the experimental conditions including the beam composition 
%and therefore the available mass calibrants, 
%and improvements made to the tune of the MRTOF system that significantly improved measurement precision.
%\emph{Experiment \#1} was performed by impinging a proton beam at an energy of 480 MeV from TRIUMF's main driver cyclotron upon a high-power tantalum target. During this experiment, yields of Yb isotopes were enhanced through resonant laser ionization.
%\emph{Experiment \#2} did the same with a low-power tantalum target.
%The proton beam current on target ranged from approximately 25 to 50 $\mu$A throughout these measurements.
%************

\begin{figure}
\centering
\includegraphics[width=0.85\linewidth]{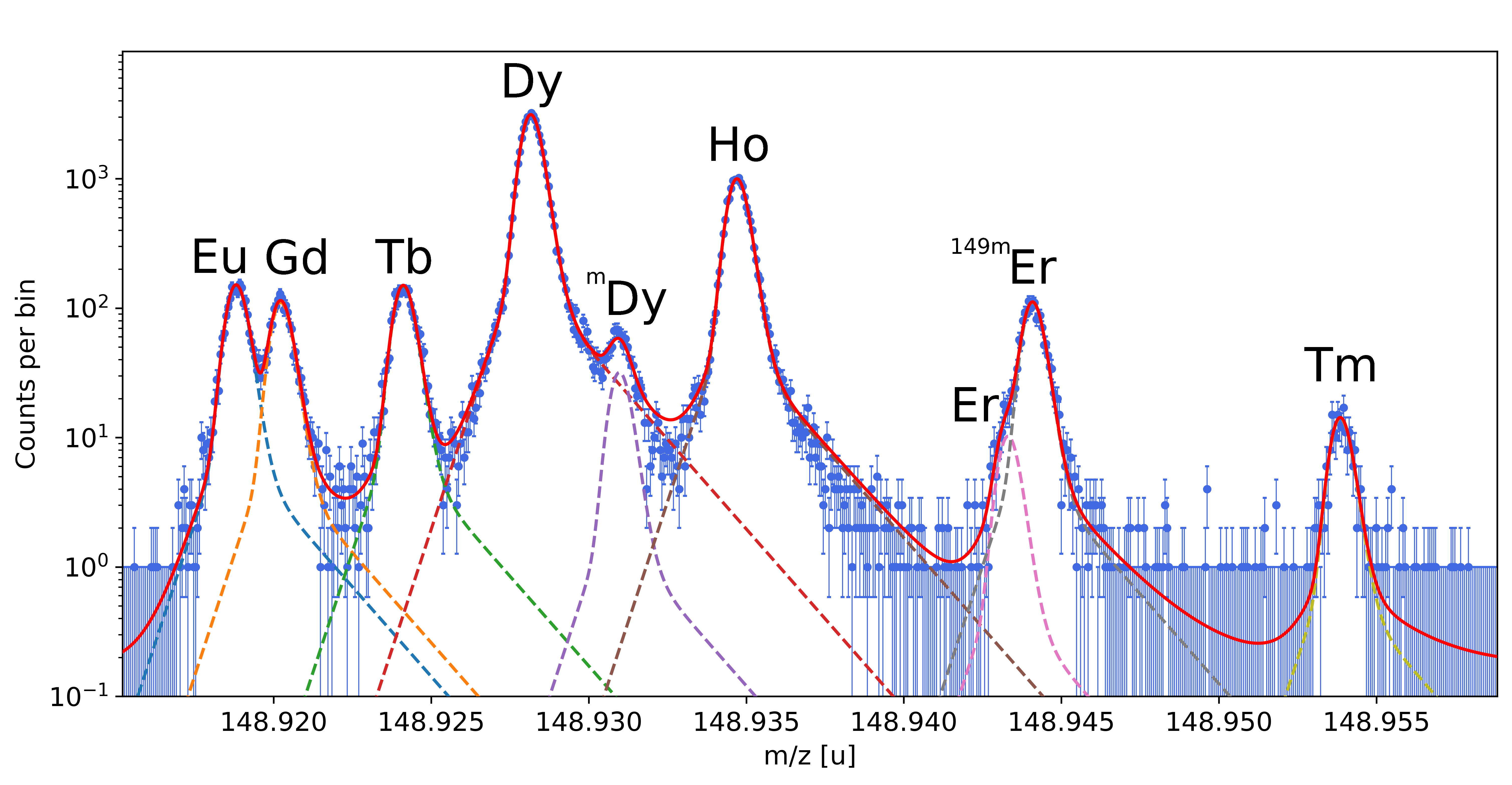} %A149Spectrum.png
%\caption[The $A=149$ spectrum with major peaks labeled]{The $A=149$ spectrum with major peaks labeled. $^{149}$Dy is used as both shape and mass calibrant here.}
%\label{fig:A149Spectrum}
\includegraphics[width=0.85\linewidth]{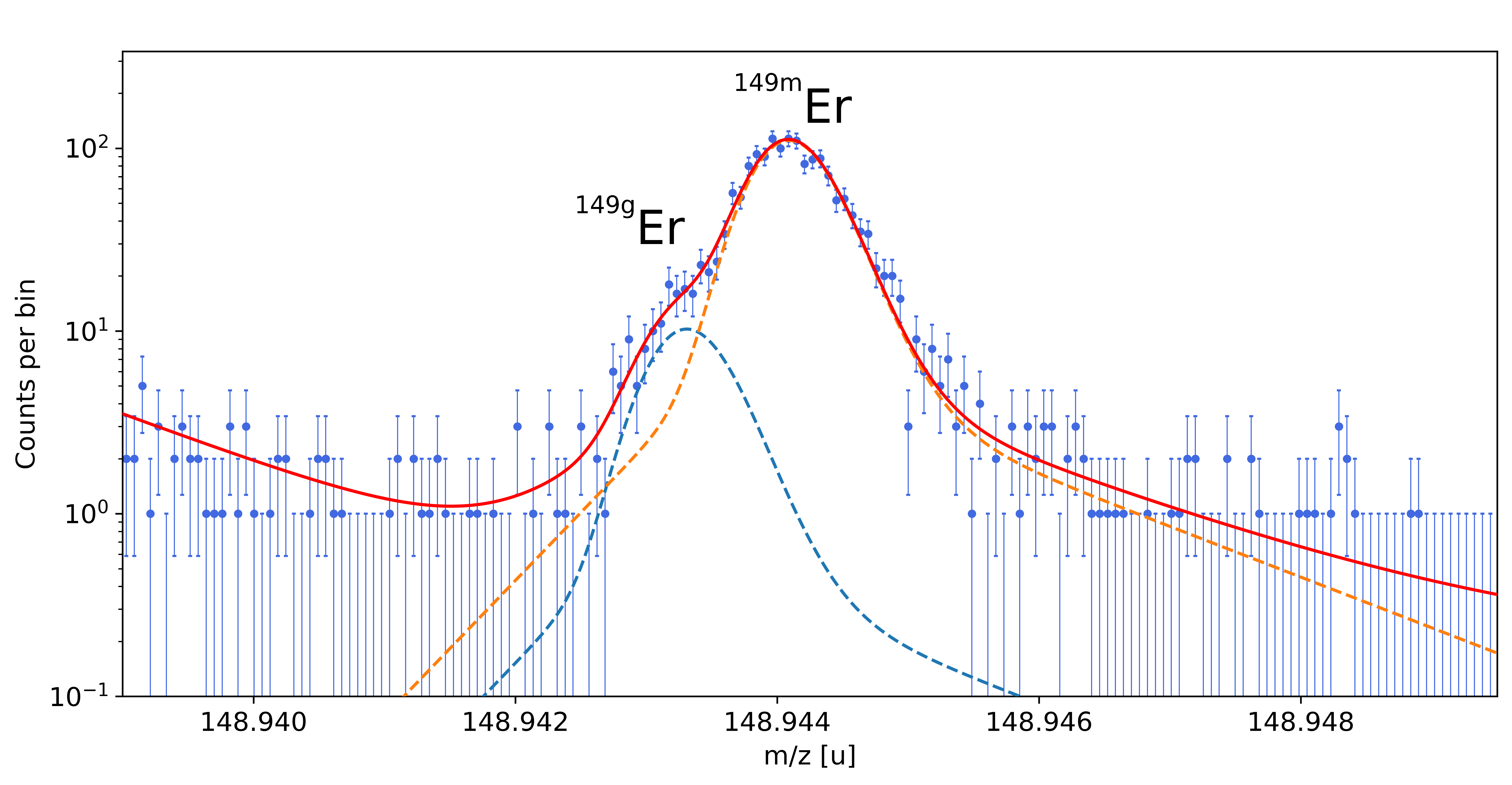}%was images/temp_149Er_pic
\caption[A fit of the $^{149g+m}$Er peak.]{Top: The $A=149$ spectrum with major peaks labeled. $^{149}$Dy is used as both shape and mass calibrant here ($m$ indicates a long-lived isomer).
Bottom: A fit of the $^{149g+m}$Er peak using the hyperEMG functional form and calibrated to the shape of the higher statistics $^{149}$Dy peak. 
The known peak shape enabled the mass of the ground state to be directly determined alongside the isomer within the mixed peak.}
%Only by accounting for the peak shape is a direct mass determination of the ground-state (contained in the left shoulder of the peak) possible.}
\label{fig:A149Spectrum}
\label{fig:149Erm_fit}
\end{figure}

The isotopes under study were produced throughout two experimental campaigns at TRIUMF's Isotope Separator and Accelerator (ISAC) facility \cite{Ball_2016} using the ISOL (isotope separation on-line) technique.
A 480 MeV proton beam was impinged on a Ta target to produce rare isotopes via spallation reactions and the produced Tm and Er isotopes were surface ionized.
The continuous beam of neutron-deficient isotopes was purified using the ISAC high-resolution mass separator \cite{BRICAULT200249} (resolving power $\approx$2500) to select beams from $A=149$ to $A = 157$ before preparation in the TITAN radio-frequency quadrupole (RFQ) cooler-buncher trap \cite{BRUNNER2012_RFQ} for precision mass measurements.
%Radioactive beams of the neutron-deficient Lanthanide elements Tm and Er and their isobars to perform precision mass measurements.
%
%During the experiment the continuous beam of isotopes from the ISAC separator was transported to the Radio Frequency Quadrupole (RFQ) cooler-buncher at TITAN and 
The resulting bunches of ions were extracted and their kinetic energy matched for injection into TITAN's MR-TOF-MS \cite{jesch2017mrTOF}\cite{dickel2019recent}\cite{REITER2021MRTOF_commissioning} to obtain a mass spectrum.
Ions were again cooled and narrowly bunched in the injection trap of the MR-TOF-MS, before they underwent isochronous (energy independent) reflections between two electrostatic ion mirrors before time-of-flight detection using a MagneTOF\textsuperscript{TM} ion detector manufactured by \emph{ETP Ion Detect}.
%\documentclass{article}
%\usepackage{textcomp}
%\begin{document}
%\textregistered\textcopyright
%\sffamily\textregistered\textcopyright
%\end{document}
All ions observed in the spectrum were delivered and measured in a 1+ charge state and any possible non-isobars present due to charge exchange were removed using a \emph{mass range selector} \cite{DICKEL2015172_MRS}. An example spectrum of $A=149$ is given in Fig.~\ref{fig:A149Spectrum}.
%The relative masses were determined via their time-of-flight at a final detector, with the mirror potentials tuned such that the ion Time-of-Flight did not depend on the ion kinetic energy for any number of reflections, and a species with a previously well-established mass was used to calibrate the spectrum.

%previously well-measured species is used to calibrate the spectrum.
%with the mirror potentials increasing the path length of the ions and maintaining their kinetic energy independence of the time-of-flight regardless of the number of reflections.

%Measurements of the species furthest from stability were made possible by utilizing \emph{Mass selective re-trapping} \cite{REITER2021MRTOF_commissioning, dickel2017isobar} to suppress the lighter isobars.

Measurements of the nuclei furthest from stability were made possible by employing so-called \emph{mass selective retrapping} \cite{DICKEL20171_retrapping, dickel2017isobar}, during which the ion bunch is captured a second time in the injection trap after a first stage of mass separation in the MR-TOF-MS. 
Mass selective retrapping can be used to selectively reduce the intensity of the dominant contaminant species by more than three orders of magnitude \cite{BeckSoenke2021_Yb}.
%to enhance the selectivity within an isobaric spectrum of radioactive beam \cite{BeckSoenke2021_Yb}\cite{REITER2021MRTOF_commissioning}.
%(due to their long half-lives, higher production rate, and/or low ionization potentials), background signals could be suppressed relative to lower-intensity exotic species.
This improved the signal-to-background ratio of the $^{149,150}$Tm measurements and allowed the MR-TOF-MS to accept a higher overall beam rate while keeping the total number of ions detected per mass measurement cycle within an acceptable range.
%(i.e. rates for which ion-ion effects did not affect the TOF in offline tests).
%...cycle below one. 

During the initial experiment (experiment 1), isotopes were produced using a high-power tantalum target which is equipped with greater heat dissipation, whereas the second experiment (experiment 2) utilized a low-power tantalum target, for which an improved and more homogeneous thermal profile is believed to have enhanced the release of Tm and Er.
In both experiments the proton beam from TRIUMF's main driver cyclotron was impinged on the Ta production targets with beam currents ranging from approximately 25 to 50 $\mu$A.
%************

%Marilena's paper: 

The mass of $^{150}$Tm was measured along with the Yb masses \cite{BeckSoenke2021_Yb} obtained in experiment 1, which additionally provided anchor masses for $\alpha$ decay chains \cite{Lykiard_pEmitters_2023}.
The resolving power of $\approx$270\,000 in this experiment resulted in a proton separation energy of $^{150}$Tm that was very close to zero, and no definitive conclusion could be drawn as to whether this nucleus is proton-bound. 
The Tm data-set was greatly improved during experiment 2 by performing additional measurements that extended data to $^{149}$Tm, and improved on the initial experimental uncertainties.
In particular, experiment 2 benefited from improvements that were made to the MR-TOF-MS system to achieve $\approx$400\,000 resolving power and lower systematic uncertainty.

%You should: Describe your methods in enough detail to allow others to evaluate and, if needed, replicate them.

\subsection{Analysis Procedure}
The analysis procedure followed the description in \cite{Ayet2019_systematics}. At all mass units, we utilized a high abundance peak to correct for time-dependent drifts in the time of flight (TOF) spectrum using in-house-developed data acquisition software \cite{dickel2019recent}.
Subsequently, masses were determined by fitting the peaks in each spectrum using a \emph{hyper-exponentially-modified Gaussian} (hyperEMG) lineshape \cite{EMG2017} and the emgfit Python library available at \cite{EMGfit}.
This procedure involves first fitting an isolated high-statistics peak with a multi-dimensional function to precisely determine the shape (i.e., functional form) of a single peak (the \emph{shape calibrant}) and then using multiple copies of this shape to fit all peaks in the spectrum.
Next, the masses of all peaks in the spectrum were calibrated against a single species with a well-established mass (the \emph{mass calibrant}).
Appropriate shape and mass calibrant peaks were chosen for each set of isobars measured to minimize the influence of overlapping isobars and nuclear isomers.
%Shape calibrants (Sept 3):
%Each mass used a common shape calibrant to determine the functional form of the hyper-EMG 
In the present analysis, the peaks corresponding to $^{149}$Dy, $^{150}$Ho, $^{151}$Ho, $^{152}$Dy, $^{153}$Dy, $^{154}$Dy, $^{155}$Yb, $^{156}$Yb, and $^{157}$Yb were selected as shape calibrants in their respective isobaric mass spectra.
The chosen mass calibrant at each mass number was cross-checked for consistency with the masses of several different isobars listed in the 2020 Atomic Mass Evaluation (AME2020) to confirm its identity, and is listed in Table~\ref{table:MassSummary}.

\subsection{Results} 

In this work, mass measurements across $17$ neutron-deficient Tm and Er isotopes were obtained, with $^{149}$Tm and $^{150}$Tm representing first-time measurements.
Mass excess values are defined as the sum of the ionic mass and the electron mass (since all ions were singly-charged) minus the mass number $A$.

The final mass values were obtained by adding the experimental and known systematic uncertainties in quadrature and making corrections for any known or expected unresolved isomers.
These are summarized in Table~\ref{table:MassSummary}.
Additionally, Table~\ref{table:MassSummary_no_isomer_correction} contains the uncorrected measurement data -- excluding any isomer corrections -- and is provided as a reference should new information about these isomers become available in the future.
Mass excess (ME) uncertainties include the uncertainty of the calibrant mass, while the ion mass ratio uncertainties do not.
The systematic uncertainty was dominated by nonideal switching of voltages in the ion mirror on the extraction side, and was measured using stable ions as described in \cite{Reiter2018} and \cite{Ayet2019_systematics}.
The single measurement presented here from experiment 1, $^{150}$Tm, includes a systematic uncertainty of $3 \times 10^{-7}$, while the measurements obtained during experiment 2 incorporate a systematic uncertainty of $1.5 \times 10^{-7}$ (except A=153). 
This reduction was the result of improvements made to the tuning of the MR-TOF-MS and the stability of the high-voltage electrodes.

\begin{figure}
\centering
\includegraphics[width=0.85\linewidth]{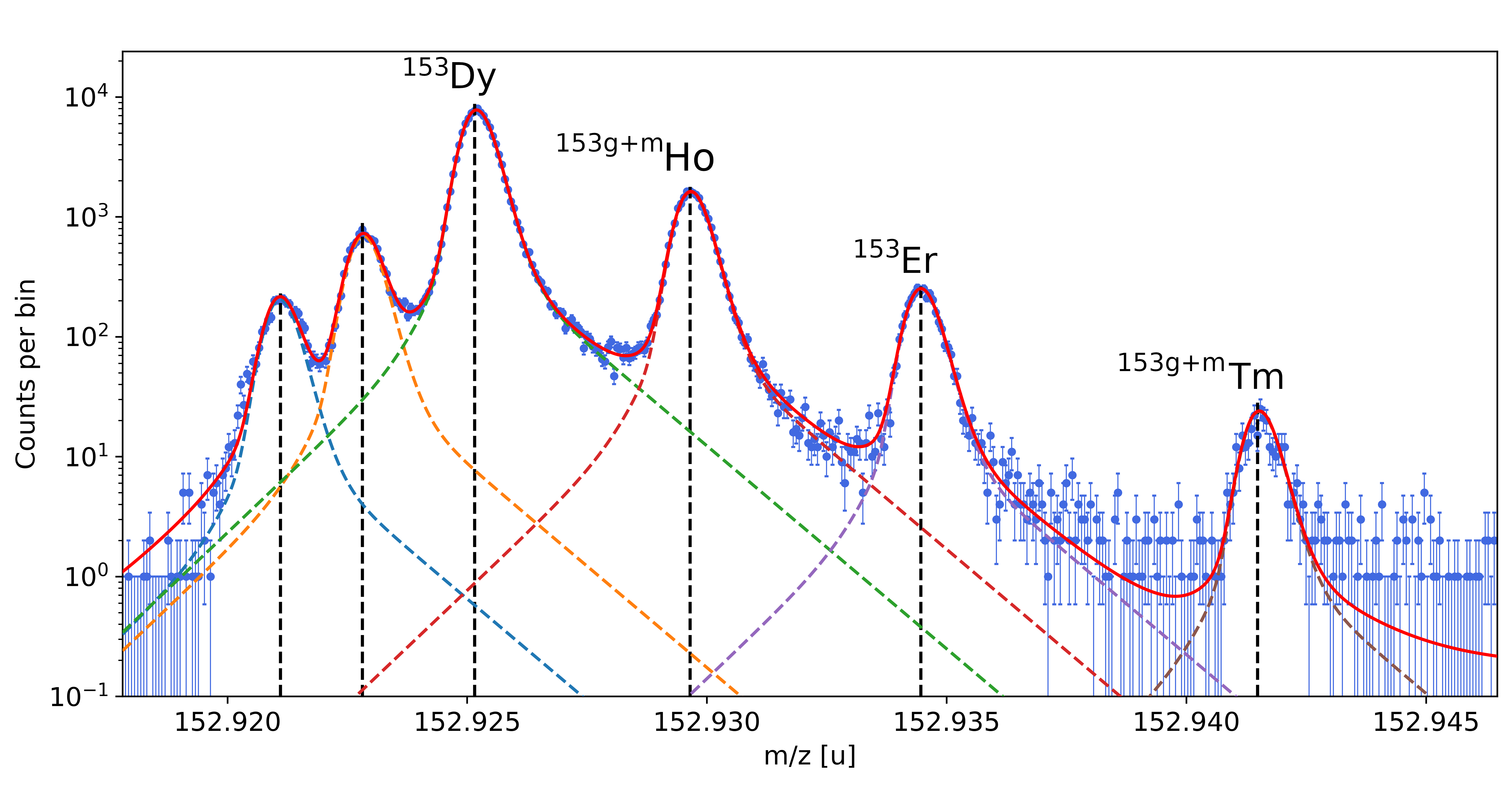} 
\caption[Fit of the $A=153$ spectrum.]{Fit of the overlapping peaks in the $A=153$ spectrum with a hyperEMG functional form.}
%Only by accounting for the peak shape is a direct mass determination of the ground-state (contained in the left shoulder of the peak) possible.}
\label{fig:A153Spectrum}
%\label{fig:149Erm_fit}
\end{figure}

%The Er isotopic chain was measured in tandem with Tm. 
%As is evident in Figure \ref{fig:Er_AME_comparison}, most values...
%Most values are consistent with literature but $^{149g}$Er is heavier by several sigma. This ground state and isomer were fit using two EMG functions calibrated to the $^{149}$Dy peak (Fig. \ref{fig:149Erm_fit}).
%the newly measured mass of $^{150}$Tm indicates that the proton separation energy ($S_p$) of the newly measured $^{150}$Tm is 
%The 150 keV excess in the mass of $^{149}$Er relative to the literature value was found in the mass spectra measuring the new Tm masses.
%
%%It is possible to explain the 136 keV deviation from literature of $^{154}$Er as resulting from contamination from an, as of yet, undiscovered isomer of $^{154}$Ho.
%It is possible that the 136 keV deviation from literature of $^{154}$Er could be influenced by contamination from an, as of yet, undiscovered isomer of $^{154}$Ho. $^{156}$Ho has both long-lived 9$^+$ and 1$^-$ isomers, $^{152}$Ho is known to have a long-lived 9$^+$ isomer which significantly impacted our Ho peak in the A=152 mass spectrum.
%152mHo (160keV)
%****
%Why is $^{153}$Er 50 keV too light? (~1.7 sigma) (154Er is ~1.3 sigma)
%****
%    \subsubsection*{$^{153}$Er}
    %$^{153}$Er was lighter than expected. 
    The crowded $A=153$ spectrum shown in Fig.~\ref{fig:A153Spectrum} was the most challenging from which to obtain a high-statistics shape calibrant in the analysis. 
    An isobaric species lies in each of the two tails of the $^{153}$Dy shape/mass calibrant, and the functional form needed to be manually selected in order to ignore excessive influence on the fitting algorithm from the isobar in the lower mass tail.
    Furthermore, the mass 153 spectrum had a total rate of $\approx$3.8 ions per trapping cycle, with $\approx$2.7 per cycle in the $^{153}$Dy calibrant peak. 
    Therefore, an additional systematic uncertainty of $8\times 10^{-8}$ per ion per cycle ($3 \times 10^{-7}$ total) was required to account for ion-ion interactions (based on off-line tests).
%
%The high rate of $^{153}$Dy also necessitated the inclusion of an additional systematic uncertainty of $3 \times 10^{-7}$. 
%    190 ions/MS at MagneTOF
%    50 Hz repetition rate during 2019 beamtime
%    135/MS in largest peak (153Dy)

%    Concern of ion-ion interactions systematically affecting the accuracy of the calibrant peak...

%\subsection*{$^{154, 152, 151}$Tm}
%The species $^{154}$Tm, $^{152}$Tm, and $^{151}$Tm each contain a long-lived isomer for which the excitation energy is unmeasured.
%Since both the excitation energy and the ratio of ground-state to isomer entering the MR-TOF-MS are unknown, it is difficult to assign a definite mass for either ground-state or isomer, but the results are included for completeness.
%****%[lower bound on isomer mass? upper bound from MRTOF resolving power? Are any double fits possible?]

%\subsection*{$^{153, 155}$Tm}\label{correction_153_155Tm}
According to ENSDF, $^{153}$Tm has a long-lived isomer at 43.2$\pm$0.2 keV \cite{Kortelahti1989_153Tm_isomer} and $^{155}$Tm one at 41$\pm$6 keV \cite{potempa1990h}.%was \cite{155Tm_isomer_etde_5590875}
%However, multiple species contained long lived isomers which were not resolvable from their respective ground-state, which may shift the resulting mass values to higher mass than expected for a pure ground-state of the given isotope.  
%When the excitation energy is known, we apply a correction and additional uncertainty following the guidelines laid out in the AME2020 appendix for the treatment of unresolved peaks.
$^{151}$Tm, $^{152}$Tm, and $^{154}$Tm %$^{151, 152, 154}$Tm %N=%_{~82, ~83, ~85} {6.6+/-2.0, 5.2+/-0.6, 3.3+/-0.07}
respectively contain isomers with half-lives $6.6 \pm 2.0$, a $5.2 \pm 0.6$, and $3.3 \pm 0.07$~s \cite{SINGH2009_ENSDF, martinNDC_2013, ENSDF_154Gd_reich2009}, for which the excitation energies were taken from NUBASE 2020 \cite{Kondev2021_nubase2020} to be 93$\pm$6, -100$\pm$250, and 70$\pm$~50 keV. %{(82, ~83, ~85)} 
NUBASE 2020 also lists extrapolated excitation energies for the unmeasured isotopes $^{149}$Tm (100$\pm$50 keV) \cite{149TmIsomer_broda1987level} and $^{150}$Tm (140$\pm$140 keV) \cite{150TmIsomer_PhysRevC.37.2694} determined from trends in neighbouring nuclei; 
the higher 671 keV, 5.2 ms isomer of $^{150}$Tm is not expected to be observable in this experiment due to its short half-life, and no indication of it was observed.
$^{157}$Lu also contains an unknown contribution from a 20.9$\pm$2.0 keV isomer (4.8 s) \cite{potempa1992investigation}.

%A corresponding additional uncertainty of 13 keV for $^{153g}$Tm and 12 keV for $^{155g}$Tm is included. %The mass of $^{155}$Tm is a mixture of the ground state and a 41 keV isomer, $^{153}$Tm the ground state and a 43.2 keV isomer, and $^{157}$Lu the ground state and a 20.9(20) keV isomer.

%possibly 151Tm isomer: \cite{151TmIsomer_2009EPJA39_49F}
%Possibly 152Tm isomer \cite{152Tm_IsomerPhysRevC.91.024322}
%Possibly 154Tm isomer: \cite{154TmIsomer_GYURKY20091}

%The mass peaks of species containing long-lived, low-lying isomers are likely an unresolved admixture of isomer and ground-state.
Since these isomeric contributions may shift the determined mass to appear heavier than for a peak containing only the ground state, the standard procedure outlined in Appendix B of AME2020 \cite{AME2020-I} was used to determine an appropriate correction to the mass and the corresponding increase in uncertainty.
The ground state mass ($m_{gs}$) was obtained from the experimental mass ($m_{exp}$) and the previously reported excitation energy ($\Delta E$) through the relation %\textcolor{red}{by subtracting half the excitation energy from the measured value, and an additional systematic uncertainty of 0.290 times the excitation energy was incorporated into the total uncertainty.}
%
%$m_{\textrm{ground state}} = m_\textrm{experiment} - \Delta E/2 $
%
%\[
$m_{\textrm{gs}} = m_\textrm{exp} - \Delta E/2$,
%\]
and additional uncertainties of $0.290(\Delta E)$ and $\frac{1}{2} \sigma_1$ were added in quadrature with the experimental error, where $\sigma_1$ is the uncertainty in the reported excitation energy.
The mass excesses and ion mass ratio were thus corrected for all species likely containing an unresolved isomer (e.g., 13 keV for $^{153g}$Tm and 12 keV for $^{155g}$Tm).
%and a corresponding additional uncertainty ($\delta m_\textrm{admixture}$) was included, where
%%
%%$\delta m_\textrm{admixture} = 0.290(\Delta E)$.
%
%for a known excitation energy $\Delta E$.
%for the treatment of overlapping peaks we follow the proceuder layed out in Appendix B of the AME 2020 part I \cite{AME2020-I}

%Either
%\subsection{$^{157}$Lu}
%or
%\subsection*{$^{157}$Lu}

%However, since only the A=157 isotope of Lu was seen, the identity of this isotope can't be confirmed. 
%The observed mass may correspond to an unidentified isobaric molecule or isomer.
%\twocolumn

The mass spectra of $^{149,150}$Tm strongly benefited from using mass selective retrapping to improve the Tm-to-contaminant ratio by about three to four orders of magnitude depending on their separation, as shown in Fig.~1 of Ref.~\cite{BeckSoenke2021_Yb}. 
Once applied, the peaks were observable above the background and could be fit.   
%While influence from $^{150m}$Tm is possible, 
%No indication of the 670 keV (5 ms) isomer of $^{150}$Tm was seen in the mass spectrum.

\begin{figure}
\centering
\includegraphics[width=0.85\linewidth]{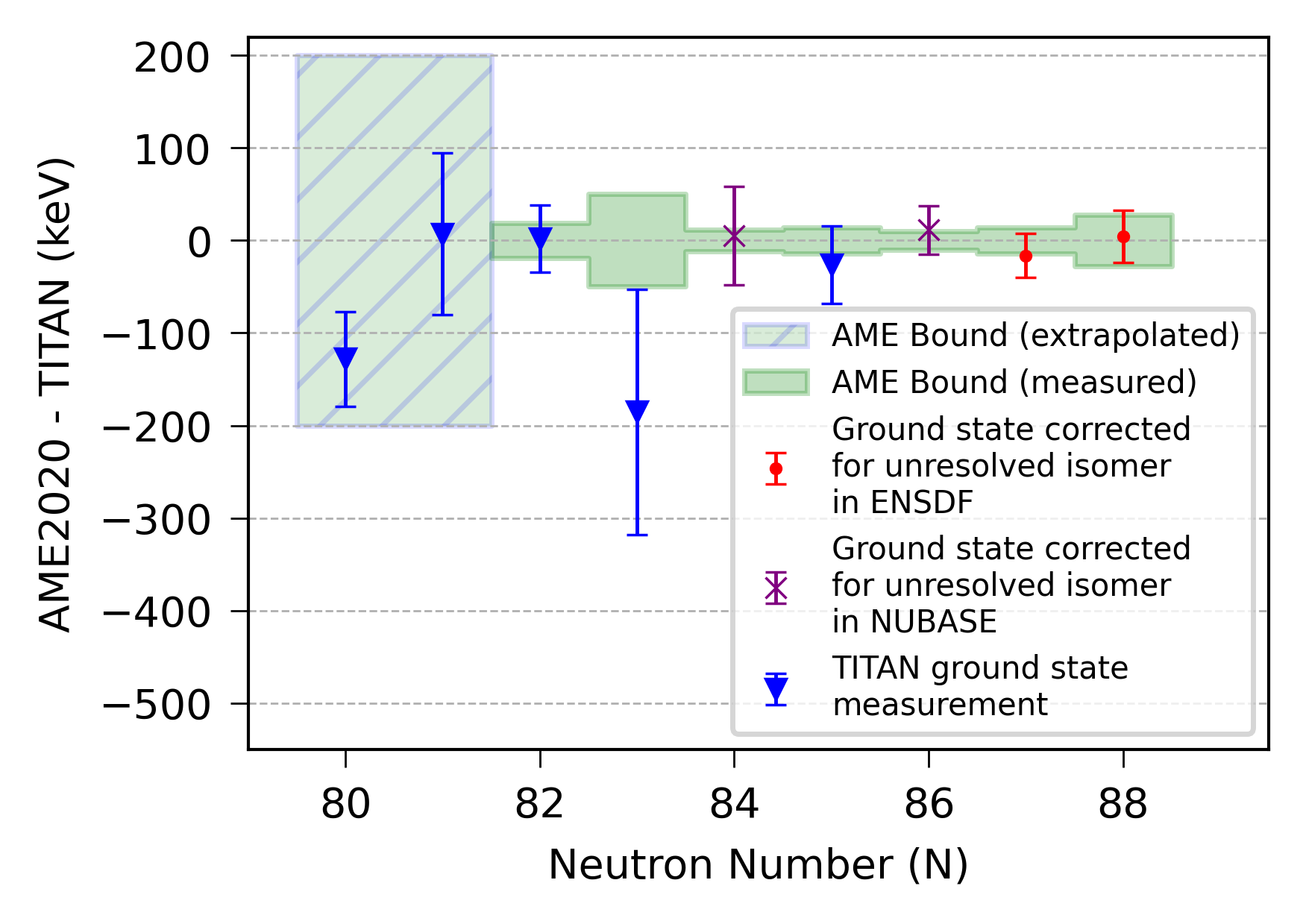}
%Tm_AME_Comparison_Apr2025.png}

\caption[]{Comparison of the measured Tm isotopes with the error bounds on the ground state given in the 2020 Atomic Mass Evaluation. The masses and error bars of measurements marked with a purple ``$\times$" or a blue triangle were corrected for unresolved long-lived isomers listed in ENSDF and NUBASE respectively.}
%Only by accounting for the peak shape is a direct mass determination of the ground-state (contained in the left shoulder of the peak) possible.}
\label{fig:Tm_AME_Comparison}
%\label{fig:149Erm_fit}
\end{figure}

As seen in Figure \ref{fig:Tm_AME_Comparison}, mass values show overall agreement with the literature values from AME2020, including those for $^{157}$Tm and $^{156}$Tm, which are not known to have long-lived isomers or isobaric species that could here impact them. 
The two new mass values of $^{149,150}$Tm now close a gap in the long chain of previously measured neutron-deficient isotopes of Tm \cite{Rauth2008_Penning_beyond_p_drip, rauth2007direct, Block2007}, with mass data now spanning from the proton emitter $^{147}$Tm at $N=78$ to stable $^{169}$Tm at $N=100$. 
Alongside the measured Tm isotopes, $^{149-154}$Er were also identified in the mass spectra.
%In addition to the mass of $^{150}$Tm obtained during the initial experiment and improved and extended to $^{149}$Tm during a second experiment, $^{149-154}$Er isotopes were also identified in the mass spectra.
The masses of $^{150-154}$Er all lie within 1.5$\sigma$ of the presently accepted values in the AME2020, and any known isomers are $> 2.5$ MeV above the ground state and thus do not impact the measurements.
After correcting for the isomer, our measurement of $^{157}$Lu yields a mass 56$\pm$29 keV heavier than the AME value. 

$^{149g}$Er and $^{149m}$Er (bottom Fig.~\ref{fig:149Erm_fit}) were fit using two hyperEMG functions calibrated to the shape of the $^{149}$Dy peak. 
The known $11/2^-$ isomeric state was delivered to TITAN at greater intensity and the ground state peak shows up as an excess in the left shoulder. 
Our $^{149g}$Er and $^{149m}$Er masses were found to be heavier by 2.9$\sigma$ and 3.9$\sigma$ respectively when compared to the literature value obtained via a measurement of $^{149m}$Er at the GSI ESR \cite{2005Li24}. 
This results in a shift of the $S_p$ of $^{150}$Tm (see below).

%The Er isotopic chain was measured in the same mass spectra as Tm and the masses of $^{150-154}$Er agree with the presently accepted values listed in the AME2020. Any known isomers are sufficiently well-separated as to not interfere with the measurement.

\begin{table*}[!ht]
\centering
\caption{List of isotopes, their respective calibrants (AME2020 values), measured ion mass ratios, and the mass excesses when referenced to the AME2020 values of the calibrant (uncertainty added in quadrature).
The $^{150}$Tm mass from experiment 1 \cite{BeckSoenke2021_Yb} is included at the bottom.
%The $^{150}$Tm mass averages data during a prior experiment \cite{BeckSoenke2021_Yb} as well as the current one. 
 The \# indicates an extrapolated mass in the AME2020. 
Nuclides with a {\bf *} have a long-lived isomer for which the excitation energy is reported in NUBASE and relies on $\alpha$ decay data, and nuclides with a {\bf **} rely on extrapolations.
Nuclides with a {\bf \textdagger} contained an unresolved admixture of the ground state and an isomer in the peak for which the known excitation energy is listed in both ENSDF and NUBASE.
The mass excess and the ion mass ratio of any peaks likely to contain an unresolved isomeric contribution were corrected by 0.5 $\times$ the listed excitation energy (listed in {\bf bold}) and a corresponding additional uncertainty was incorporated (see Sec.~\ref{sec:Discussion}).
%; the mass excess and uncertainty were corrected to account for this. 
The $^{155}$Tm peak has an unresolved 41(6) keV isomer \cite{potempa1990h}, $^{153}$Tm has a 43.2$\pm$0.2 keV isomer \cite{Kortelahti1989_153Tm_isomer}, and $^{157}$Lu has a 20.9$\pm$2.0 keV isomer \cite{potempa1992investigation}. 
As described in the text, these values were corrected using the procedure outlined in the 2020 Atomic Mass Evaluation (AME2020) \cite{AME2020-I} to obtain a mass for the ground state and its uncertainty. 
The difference between the TITAN measurement and the AME2020 value is given with the AME and the measurement uncertainty added in quadrature.
}
\small % Adjust the font size to be smaller
\begin{tabular*}{\textwidth}{@{\extracolsep{\fill}} c c c c c c}

\hline \hline \\ [-1em]
Species & Calibrant & Ion Mass Ratio & ME$_{\textrm{Measured}}$ (keV) & ME$_{\textrm{AME2020}}$ (keV) & TITAN -- AME (keV)\\
[-1em]\\
\hline \\ [-1em]
%First attempt at mass ratio was 1.000048202
{\bf $^{157g}$Lu {\bf \textdagger}} & $^{157}$Yb & {\bf 1.00004813(16)} & {\bf -46384 (26)} & -46440 (12) & 56(29) \\ % mass form fit: -46373.34 keV %20.9(20) isomer
$^{157}$Tm & $^{157}$Yb & 0.99996379(18) & -58714 (28) & -58709 (28) & -4(40) \\ 
$^{156}$Tm & $^{156}$Yb & 0.99997554(15) & -56818 (24) & -56834 (14) & 17(28) \\ 
{\bf $^{155g}$Tm {\bf \textdagger}} & $^{155}$Er & {\bf 1.00003875(17)} & {\bf -56617 (26)} & -56626 (10) & 9(28) \\ 
$^{155}$Yb & $^{155}$Er & 1.00008157(15) & -50437 (23) & -50503 (17) & 66(28) \\ %diff was  & 66(29)
$^{154g}$Tm {\bf *} & $^{154}$Dy & {\bf 1.00011154(29)} & {\bf -54401 (42)} & -54427 (14) & 26(44) \\ %IMR was 1.00011179(18) ME was -54366 (27)  %diff was & 61(30)
$^{154}$Er & $^{154}$Dy & 1.00005458(16) & -62569 (24) & -62605 (5) & 36(24) \\ %diff was & 36(25)
{\bf $^{153g}$Tm {\bf \textdagger}} & $^{153}$Dy & {\bf 1.00010661(37)} & {\bf -53957 (54)} & -53973 (12) & 16(55) \\ %was {\bf 1.00010662(38)} & {\bf -53957 (54)} % recently ME was changed from -53955 (55) to -53957 (54) % was -53973 {\bf (34)} giving a difference of 0
$^{153}$Er & $^{153}$Dy & 1.00006064(34) & -60505 (48) & -60467 (9) & -39(49) \\%was 1.00006064(34) & -60505 (48) % diff was & -50(49)
$^{152g}$Tm {\bf *} & $^{152}$Dy & {\bf 1.00013131(94)} & {\bf -51535 (133)} & -51720 (50) & 186(142) \\ %IMR was 1.00013096(23) & {\it -51585 (33)}  %diff was & 136(60)
$^{152}$Er & $^{152}$Dy & 1.00006789(18) & -60510 (26) & -60500 (9) & -10(27) \\ %
$^{151g}$Tm {\bf *} & $^{151}$Dy & {\bf 1.00012788(26)} & {\bf -50774 (36)} & -50772 (19) & -3(41) \\ %IMR was 1.00012821(17) & {\it -50728 (24)}     %Diff was & 44(31)
$^{151}$Er & $^{151}$Dy & 1.00007472(15) & -58248 (22) & -58266 (16) & 19(27) \\ 
$^{150g}$Tm {\bf **} & $^{150}$Dy & {\bf 1.00016335(62)} & {\bf -46497 (87)} & -46490 (200\#) & \#\# \\ %64 %%Before isomer correction was 1.00016386(23) & -46427 (33)
$^{150}$Er & $^{150}$Dy & 1.00008237(16) & -57807 (23) & -57831 (17) & 24(29) \\ 
$^{149g}$Tm {\bf **} & $^{149}$Dy & {\bf 1.00017215(36)} & {\bf -43812 (51)} & -43940 (200\#) & \#\# \\ %179  %was 1.00017215(36) & -43762 (51)
$^{149g}$Er & $^{149}$Dy & 1.00010171(33) & -53584 (47) & -53742 (28) & 158(55) \\ 
$^{149m}$Er & $^{149}$Dy & 1.00010695(16) & -52857 (25) & -53000 (28) & 143(37) \\%was 143(38) 
$^{150g}$Tm (Expt. 1){\bf **} & $^{150}$Dy & {\bf 1.00016357(73)} & {\bf -46467 (103)} & -46490 (200\#) & \#\# \\ %93  %% Before isomer correction this was 1.00016407(34) & -46397 (47)
\hline \hline
\end{tabular*}

\label{table:MassSummary}
\end{table*}

\renewcommand{\arraystretch}{1.2} % Increases the row height by 20%

\section{Discussion}
\label{sec:Discussion}

%isotope chains have been used to pin down the Tm drip-line and test the N=82 shell closure.
As detailed below, our measurements of the Tm and Er mass chains establish $^{149}$Tm as the first proton-unbound Tm isotope.
%The Tm and Er mass chains we measured establish that $^{149}$Tm is the first Tm isotope capable of proton emission.
%and $^{149-154}$Er 
Furthermore, they support recent evidence for the persistence of the shell closure at $Z=70$ \cite{BeckSoenke2021_Yb} with a corresponding measurement of the trend in $\Delta_{2N}$ at $Z=69$.

\subsection*{Last Proton-Bound Isotope Determined from $^{149m, 149g}$Er and $^{150g}$Tm}%{149gEr}

%Since the Er isobars tend to be more stable than neutron-deficient Tm, several Er isotopes could be measured in tandem during the second experiment ($^{154}$Er to $^{149}$Er).
%
%	Neutron-deficient Er being one element closer to stability than Tm, the Er isotopic chain...
The Er isotopic chain was measured alongside Tm during experiment 2 from $^{154}$Er to $^{149}$Er. 
%    
%	Good agreement was found with the accepted values of the masses of the Er isotopes, with the exception of $^{149}$Er.
 %
%	The masses of $^{149g}$Er and its corresponding isomer, $^{149m}$Er lie  $\approx$ 150 $\pm$ 29 keV/c$^2$ higher than the literature value \cite{AME2020-II}.
%
During this second measurement, a discrepancy in the mass of $^{149g}$Er relative to the literature value was found that significantly affects the one-proton separation energy ($S_p$) of the $^{150}$Tm ground state.
Together, the newly measured $^{150}$Tm mass and the AME2020 mass of $^{149g}$Er results in a $S_p$ that is well within $1\sigma$ of zero, but using both updated masses yields a positive proton separation energy consistent with proton-bound $^{150}$Tm ($2\sigma$).
%Without taking into account this adjustment to the mass of $^{149g}$Er the measured proton separation energy ($S_p$) of $^{150}$Tm is consistent with zero ($1\sigma$), but when this updated mass of $^{149}$Er is included the proton separation energy is determined to be positive, resulting in $^{150}$Tm being the last proton-bound isotope.
%%and therefore the isotope lies on the proton-bound side of the drip-line.
%Masses of both the parent and progeny nuclei allows the 1-proton separation energy of $^{150}$Tm to be calculated directly without any dependence on prior measurements (aside from the masses of the proton and the electron), and establishes $^{150}$Tm to be the last proton-bound Tm isotope.

%Similar to the $^{151}$Yb isotope (also at N=81) which was observed in our prior work \cite{BeckSoenke2021_Yb}, 
As with $^{151}$Yb, for which the ISAC target was seen to preferentially deliver the spin 11/2 isomer over the ground state, $^{149}$Er, having two fewer protons, similarly appeared to favour the excited state.
%Since the ISAC target was seen to preferentially deliver the spin 11/2 isomer of $^{151}$Yb to the ground-state, a similar relative abundance is expected for its isotone $^{149}$Er, having two fewer protons.
We therefore fit the two-component $^{149}$Er peak using two hyperEMG functions having the same parameters as the $^{149}$Dy peak; this yields mass values for both the dominant isomeric state ($t_{1/2} = 8.9\pm$0.2 s), as well as the ground state ($t_{1/2} = 4\pm$2 s) which shows up as an excess within the low-mass tail of the isomer peak. 
%
%($^{149m}$Er having a half-life of 8.9$\pm$0.2 s vs $^{149g}$Er with a half-life of 4$\pm$2 s),
%150Tm has 671.3 +/-10keV isomer

%As such, t
The ground state mass of $^{149}$Er can be determined by two different methods using our data.
One approach is to extract it from the measured isomer mass and the known excitation energy as was previously done in an experiment using Schottky mass spectrometry 
%at the FRS-ESR storage ring at GSI 
\cite{2005Li24}, which reported an uncertainty of $\pm$28 keV.
The other method is to obtain the mass directly from the fit of the ground state within the tail of the isomer using the known peak shape.% from $^{149}$Dy.

%The ground state was delivered to our experiment with lower abundance, but shows up as an excess within the low-mass tail of the $^{149m}$Er isomeric peak.

%This increases our confidence that the peak is dominated by the isomeric state.

The first approach obtains the mass of $^{149g}$Er using the known isomeric excitation energy, which was initially measured via a two $\gamma$-ray internal transition to the ground state proceeding through a 630.5 keV (\textit{M}4) $\rightarrow$ 111.0 keV (\textit{M}1) cascade \cite{Toth1985}  
This was later confirmed in subsequent experiments which yielded 630.3 keV $\rightarrow$ 111.3 keV \cite{149TmIsomer_broda1987level} and 630.5$\pm$2.6 keV $\rightarrow$ 111.3$\pm$1.1 keV by \cite{Firestone1989}. %\cite{Firestone1989}.
The $^{149m}$Er isomer is established to lie 741.8$\pm$0.2 keV above the ground state according to the NUBASE 2020 evaluation \cite{Kondev2021_nubase2020}. 
%The decay energy was previously determined through gamma ray spectroscopy experiments, the isomer being observed to transition to the ground state through a  \cite{Toth1985}which was later confirmed through a measurement by \cite{broda1987level} of 630.3 keV 111.3 keV and a measurement of 630.5 (2.6) keV 111.3 (1.1) keV by \cite{Firestone1989}.

%Furthermore, as discussed in the recent paper detailing results from the measurement campaign in which the mass of $^{150}$Tm was initially obtained%experiment \#1 data-set, the isomeric chain containing $^{149}$Er has been shown to exhibit a remarkably consistent excitation energy of approximately 750 keV, which has now also been observed at Z=70 \cite{BeckSoenke2021_Yb}.
This excitation energy is consistent with our recent result in which the newly measured masses of $^{151g+m}$Yb %experiment \#1 data-set
were used to support a theoretical explanation for the remarkably consistent first excitation energy (approximately 750 keV) in the isotonic chain containing $^{149}$Er \cite{BeckSoenke2021_Yb}.
%; $^{149}$Tm is now bracketed by measurements at Z=70 and 68 downwards.
This has been attributed to a clustering of single-proton energy levels for the mildly deformed nuclei at $N=81$ \cite{BeckSoenke2021_Yb} resulting from the single hole in the $h_{11/2}$ neutron orbital common to $^{149}$Er and many of its isotones \cite{Toth1985}.
This provides increased confidence in the reliability of the established excitation energy of the long-lived 741.69$\pm$0.23 keV isomer $^{149m}$Er (9.6$\pm$0.6 s).
%\cite{AME2020-II}
Using this value with our current mass measurement of $^{149m}$Er to indirectly determine the ground state mass yields a $^{149g}$Er mass excess of -53599$\pm$25 keV, a 143 keV increase
%-53,598.8(13)(21)(0.2) keV, a 143 keV/c$^2$ increase
from AME2020.
%(-52857(13)() - 741.8$\pm$0.2 keV)
% (from: 741.69 +/-23)
%
%was% ...(9.6$\pm$0.6 s), from the value determined using the presently
%An increase of 143 $\pm$ 25 keV/c$^2$ in the mass of $^{149g}$Er
%This implies an increase of 143 keV/c$^2$ in the mass of $^{149g}$Er when using the mass of the long-lived 741.69 $\pm$ .23 keV isomer (9.6$\pm$.6 s) and its presently accepted excitation energy \cite{AME2020-II} instead of directly measuring the ground state. %was% 150$\pm$29 keV

However, the present dataset also enables a more direct measurement of the ground state mass by utilizing the lower abundance, overlapping TOF peak of the ground state of $^{149}$Er and employing a hyperEMG fitting procedure. %a \emph{hyper Exponentially Modified Gaussian} fit (hyperEMG). 
%This was valuable since the Gaussian approach could not directly determine the mass of the $^{149}$Er ground state.
Directly fitting the lower abundance $^{149g}$Er (as in Fig.~\ref{fig:149Erm_fit}) with the hyperEMG function calibrated to the shape of the $^{149}$Dy peak yields a mass excess of $\text{-53584$\pm$47}$ keV, an increase in mass of 158$\pm$55 keV from the AME2020 value.
The excitation energy of $^{149m}$Er can then be determined from the difference between our measurements of ground state and isomer to be 726$\pm$53~keV.%$726(44)(30)$~keV.
% was: \emph{726(44)()() keV} 727?.
%Between Soenke's value of 680 keV and the trend of ~750 keV below it
%This amounts to 158 $\pm$ 47 keV/c$^2$ increase from the AME2020 value (28 keV uncertainty), 
%The direct fit yields an increase in mass of 158(55) of the ground state from the AME2020 value.
%The direct fit yields an increase in mass of 158 $\pm$ 47 keV/c$^2$ of the ground state from the AME2020 value.
%from the AME2020 value (28 keV uncertainty), and allows us to positively establish the proton-bound nature of $^{150}$Tm.
%This method extracted the mass to be 158 $\pm$ 47 keV/c$^2$ heavier than the AME value which itself has an uncertainty of 28 keV.
%This provides an updated ground state mass of $^{149}$Er that allows us to establish the proton-bound nature of $^{150}$Tm.

%Move to conclusion???
    While the newly determined mass of $^{149g}$Er and the isomer $^{149m}$Er 
%    While the newly determined mass of $^{149g}$Er shown in Figure \ref{fig:Er_AME_comparison} and the isomer $^{149m}$Er 
    %are in tension with
    deviate from the mass value in the AME2020, they are consistent with observable trends in the binding energy of nuclides in this region%such as the empirical shell gaps for proton and neutron binding ($S_p$ and $S_n$) and the two-neutron separation energy
    .
The cause of our discrepancy in our measurement of this single mass is unknown, but may warrant a third measurement of the $^{149g,m}$Er isotope for confirmation.

%Ultimately, this new measurement of the isomer leads us to the conclusion that an updated mass is needed for the ground $^{149}$Er state.
%Although the cause of the discrepancy from literature is unknown, our confidence in the reliability of the ground state excitation energy combined with the established accuracy of other masses determined using the MRTOF, leads us to adopt a revised value of $^{149g}$Er from our measurement of the $^{149m}$Er (11/2-) isomer.
These new masses of the $^{149}$Er isomer and ground state lead us to adopt a revised one-proton separation energy of the ground state, which 
%value for both ground state and isomer of $^{149}$Er.
%This shift in the $^{149}$Er mass 
directly impacts the location of the proton drip-line in Tm, since the proton separation energy of a nucleus depends on both the mass of the candidate proton emitter as well as on the mass of the daughter nucleus of the decay.
Both the $^{149,150}$Tm masses as well as the revised $^{149}$Er value were thus needed for an unambiguous identification of the Tm drip-line, since the new Tm masses alone did not exclude the possibility that $^{150}$Tm could also be proton-unbound when using the $^{149}$Er value from the AME2020.
These measurements confirm that $^{150}$Tm is indeed proton-bound, 
%the most neutron-deficient proton-bound ground state in the Z=69 isotopic chain.
%
%Furthermore, a bound $^{149}$Tm would be at odds with the predictions of \ref{Tm_drip_2020}.
%The drip-line location determined here agrees
agreeing qualitatively with the \emph{Bayesian model averaging} predictions  of \cite{Tm_drip_2020} that $^{149}$Tm is the first proton-unbound thulium isotope.
Furthermore, the proton emission $Q$-value of $^{149}$Tm has been determined and can now be used as an input for decay rate calculations.
%(where $Q_p = -S_p = M(A,Z) - M(A-1,Z-1) - $M$(^{1}$H$)$).

%***
%Q-value -132(25)(21),(42)(21) [What happens to the Q-value systematic error when we measure both 149Er and 150Tm the same way?]
%***
%***

%Influence from the $^{150}$Tm isomer on this measurement is expected to be minimal since the ground state has a much longer half-life. Furthermore, it would only serve to inflate the value of $S_p$ and increase the binding of the ground state.

%Isotopic trends from adjacent mass units were used to exclude the possibility that a molecule of very similar mass could have impacted the measurement.

%  Isotopic trends from adjacent mass units were used to exclude the possibility that a molecule of very similar mass could have impacted the measurement, and there is no known higher-lying isomer of $^{149}$Er that could have been mistakenly measured during the storage ring experiment that could explain the tension with our present measurement.

\subsection*{$^{149}$Tm: The First p-Unbound Isotope}

The proton emission energy can be determined from \( Q = -S_p = M(Z,N) - M(Z-1,N) - M({}^1\mathrm{H}) \). 
%\(Q = M(Z,N) - M(Z-1,N) - M($^1H)\).
Our direct measurements of $^{149g}$Er and $^{150g}$Tm show $^{150g}$Tm to be proton-bound, having a $Q$-value of -202$\pm$99 keV.
Meanwhile, our measurement of $^{149g}$Tm establishes it as the first proton-unbound isotope in the thulium chain.
For it we obtain a proton emission $Q$-value of +378$\pm$52 keV when calculated using the measured mass of $^{149g}$Tm and the masses of $^{148}$Er and $^{1}$H from the AME2020. 
%Nuclear from Atomic mass:
%M_N = M_A - Z \times m_e + B_e(Z)
%Using \(Q = M_{parent} - \Sigma M_{products}\) gives %$Q_p = M_{^{149}Tm} - M_{^{148}Er} - M_p = 0.000446 \pm 0.000051$AMU $= 420 \pm 48$ keV/c$^2$
%
%$Q_p = M_{^{149}Tm} - M_{^{148}Er} - M_{^1\textrm{H}} = 428 (35) \textrm{keV}$, % was: (27)(21)(10) 
%\begin{equation*}
%\begin{split}
%Q_p &= M_{^{149}Tm} - M_{^{148}Er} - M_p,\\
%	&= 0.000446 \pm 0.000051 \textrm{AMU},\\
%	&= 428 (27)(21)(10) \textrm{keV},
    %&= 420 \pm 48 \textrm{keV/c}^2,
    %(35.6)???
%\end{split}
%\end{equation*} 
%
This decay energy is on the high side of the AME extrapolation of 250$\pm$201 keV and can be used in estimations of the proton emission half-life for a given angular momentum transfer (e.g. \cite{Delion2006PRL, Zhang2018}).
The small $Q$-value of the $^{149g}$Tm decay allows for it to proceed to only one possible final state, the 0+ ground state of $^{148}$Er, so it is expected that J$^{\pi} = 11/2^-$.
Both of the above decay $Q$-values and their uncertainties take into account the expected unresolved isomers listed in NUBASE inferred from trends in neighbouring nuclei.

\begin{figure}
\centering
\includegraphics[width=0.98\linewidth]{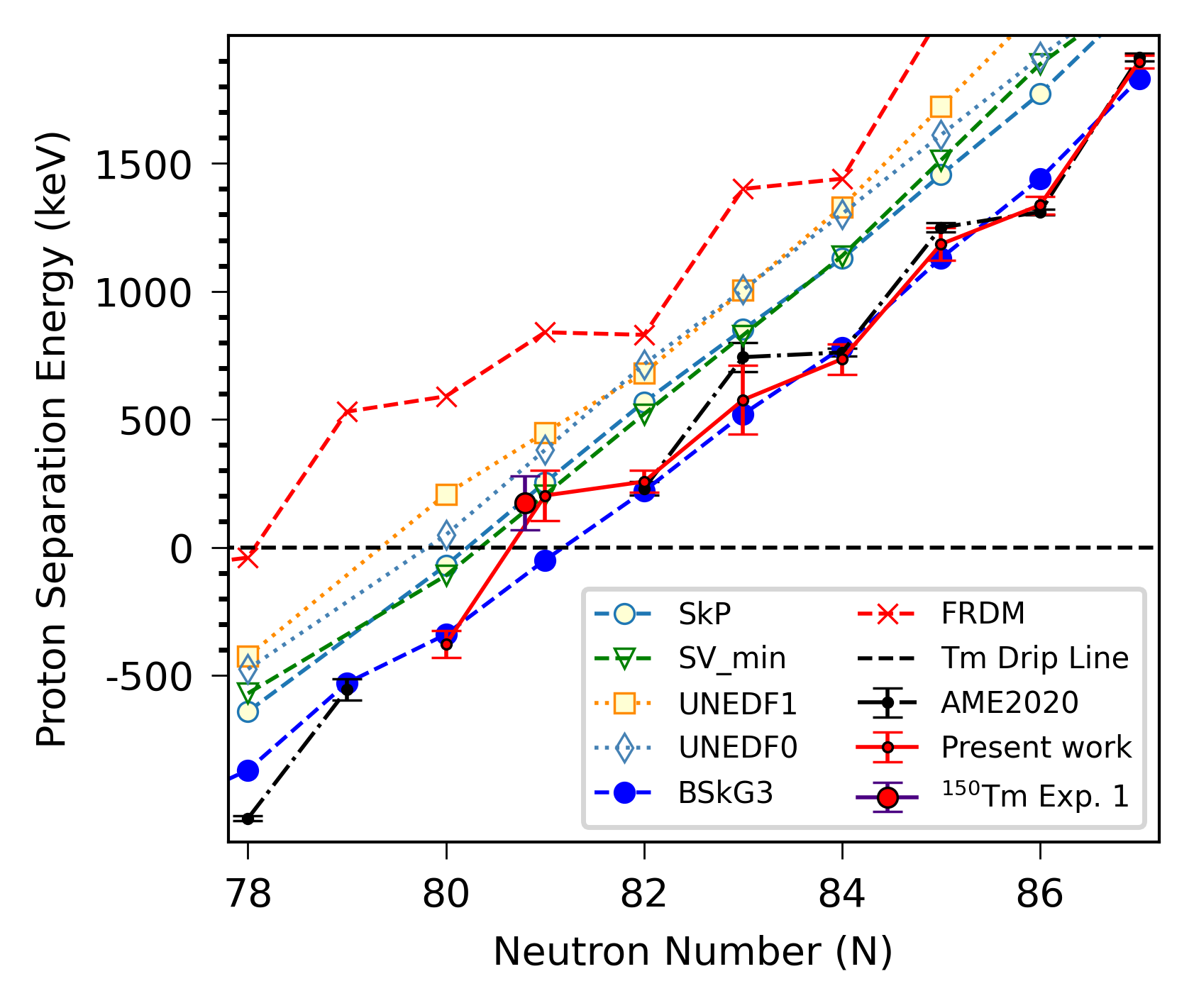}
\caption[The one-proton separation energy of Tm ($Z=69$) calculated with the newly measured Tm and Er masses.]{The one-proton separation energy ($S_{p}$) of Tm isotopes calculated from the present measurements of masses in the Tm ($Z=69$) and Er ($Z=68$) isotopic chains is shown in red and connected by solid lines (the $^{150}$Tm data point from experiment 1 is offset horizontally for clarity). This includes our new mass of $^{149}$Er at $N=81$, but masses from the 2020 AME \cite{AME2020-II} were used for $^1$H, $^{148}$Er ($N=80$), and $^{155}$Er ($N=87$) in calculating $S_{p}$ for Tm.
Theory calculations of $S_{p}$ are shown for comparison.
%Density functional theory calculations {\color{red} using the methodology of} \cite{ERLERJochen2012Tlot} are shown in dotted lines. 
Corrections were made for measurements possibly containing unresolved, long-lived, isomeric states (including the extrapolated isomers in $^{149,150}$Tm) but do not impact the conclusion that $^{150}$Tm is the last proton-bound Tm isotope.
%The red data point represents the separation energy if the $^{149}$Er mass from the 2020 AME \cite{AME2020-II} is used instead of the new measurement.
}
\label{fig:Tm_chain_1p_separation_E}
\end{figure}

Aside from BSkG3 \cite{BSkG3_grams2023skyrme}, the models in Fig.~\ref{fig:Tm_chain_1p_separation_E} systematically over predict $S_p$.
Thus the present measurements demonstrate that $^{149}$Tm is the first proton-unbound nucleus, rather than $^{148}$Tm or $^{150}$Tm.

\subsection*{Evolution of the $N=82$ Neutron Shell Closure}

\begin{figure}
\centering
\includegraphics[width=1.05\linewidth]{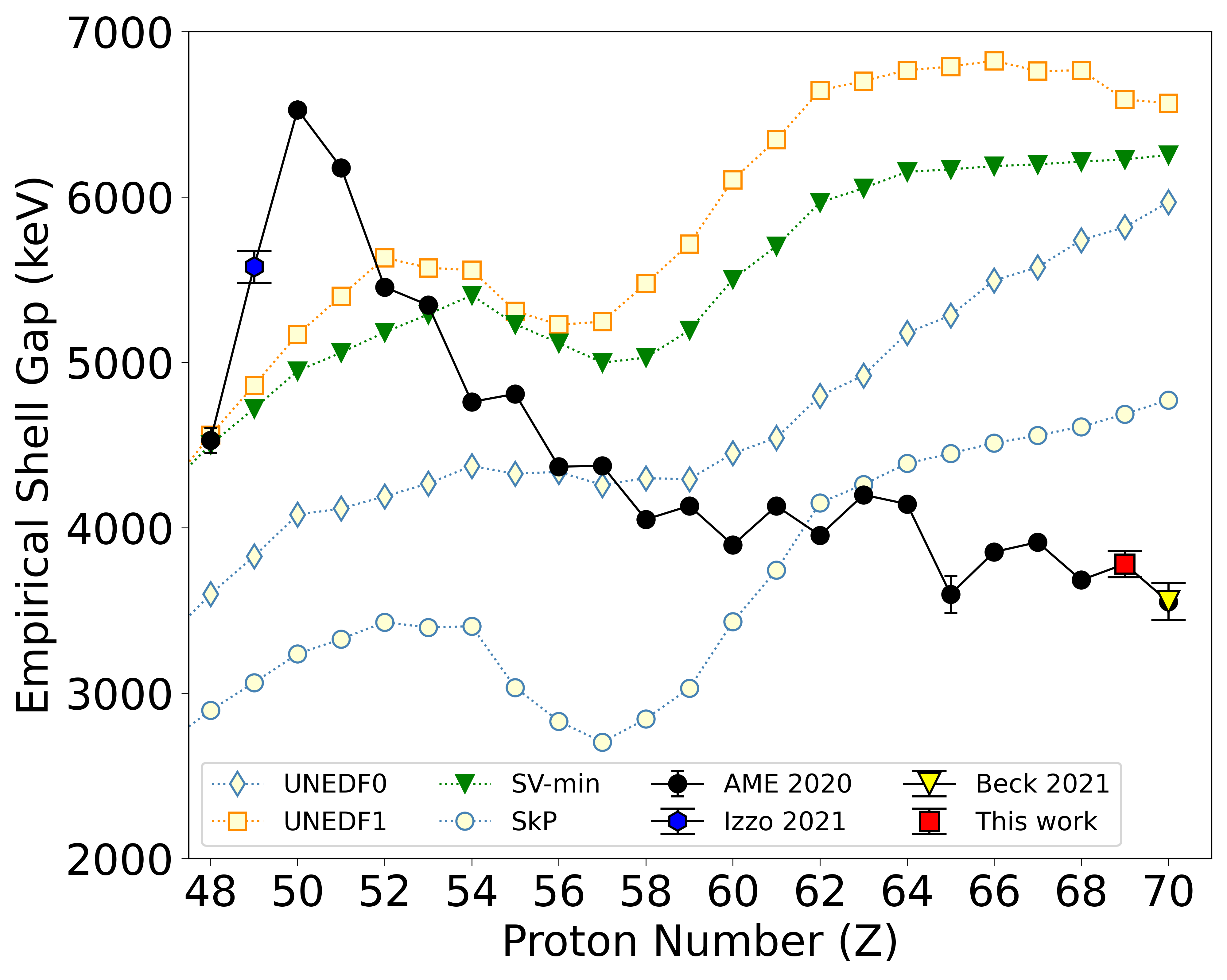}%DFT_D2N_comparison_Sept4.png
\label{fig:DFT_Comparison}
\includegraphics[width=1.05\linewidth]{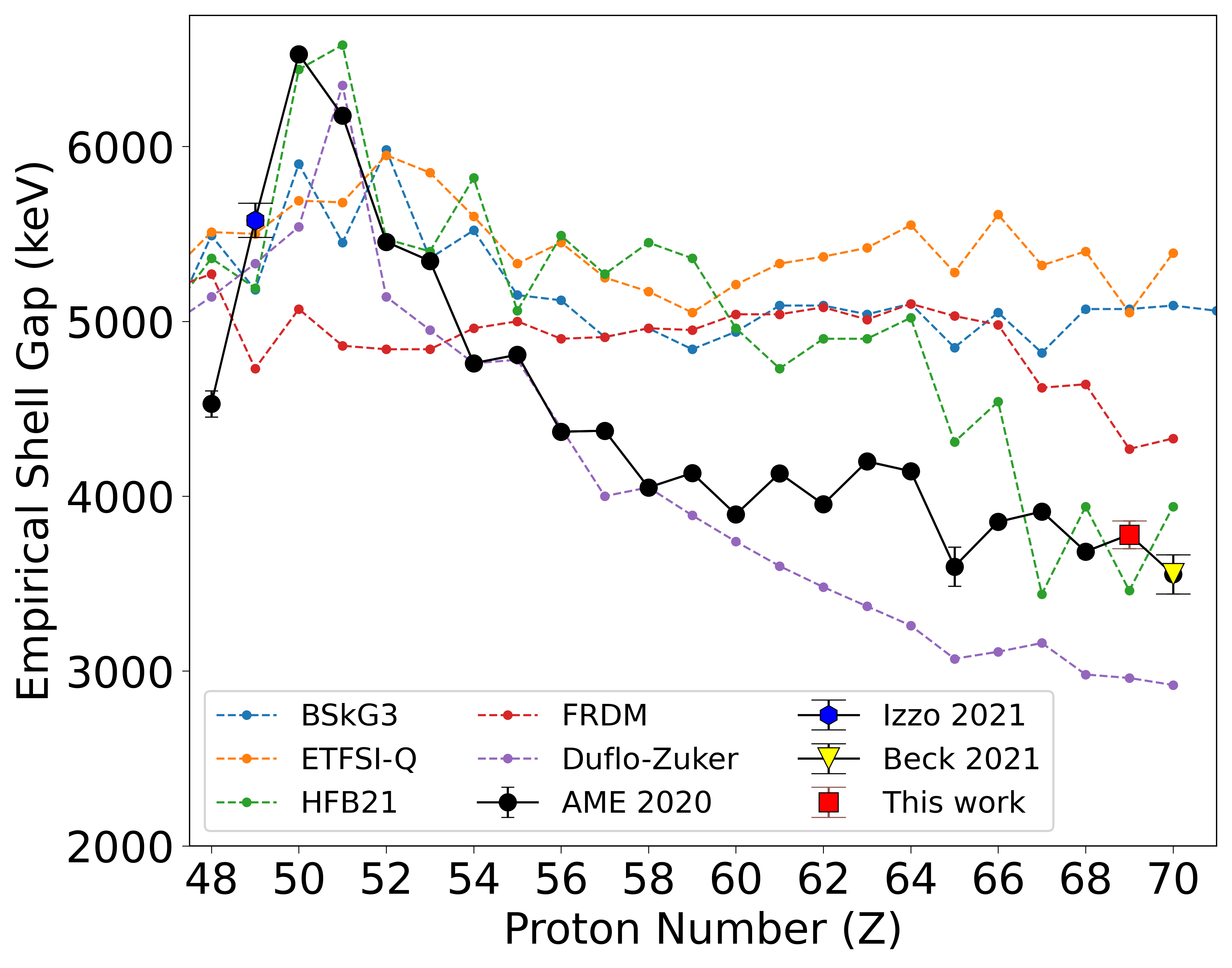}%Droplet_D2N_comparison_Sept4.png
\caption[]{Comparison of models and experimental data for the empirical shell gap ($\Delta _{2N}$) split into two figures for clarity. The mass data from \cite{Izzo2021_In} and \cite{BeckSoenke2021_Yb} was added for $Z=49$ and $Z=70$ respectively. A selection of density functional theory and ``liquid drop" models are shown.}
%\caption[]{Comparison of Density Functional Theories (DFTs). The mass data from \cite{Izzo2021_In} and \cite{BeckSoenke2021_Yb} was added for $Z=49$ and $Z=90$ respectively.}
\label{fig:Droplet_Comparison}
\end{figure}

In addition to determining the location of the drip-line, the new masses of $^{149/150}$Tm allow for a comparison to predictions of the empirical shell gap of the $N=82$ shell closure defined by $\Delta_{2N} = ME(N,Z+2) + ME(N,Z-2) - 2ME(N,Z)$ (see %Figure \ref{fig:DFT_Comparison} and 
Fig.~\ref{fig:Droplet_Comparison}).
%Many DFT models over-predict $\Delta_{2N}$ for these neutron-deficient isotones, while droplet-like models seem to predict the strength of the $N=82$ more reasonably. 
The models shown include UNEDF0 \cite{Kortelainen2010UNEDF0}, UNEDF1 \cite{Kortelainen2012UNEDF1}, HFB21 \cite{goriely2010HFB21}, Duflo-Zuker \cite{duflo-zuker1995microscopic}, FRDM \cite{moller1988FRDM}, SkP \cite{dobaczewski1984_SkP}, SV-min \cite{Klupfel2009_SV-min}, and ETFSI-Q \cite{pearson1996_ETFSI-Q}.
In this region most models over predict the shell gap of the $N=82$ isotone chain by 0.5--3 MeV with the exceptions of Duflo-Zuker under predicting by nearly 1 MeV and HFB21 which lies within 500 keV of the isotones from $Z=67$ to $70$.

%[Compare DFT models to extrapolations and give references!!]
The present work measures the $\Delta_{2N}$ of Tm at the $N=82$ shell closure and shows that it is, in fact, the first isotone for which $\Delta_{2N}$ depends on the binding energy of a proton-unbound nucleus ($^{149}$Tm).
These measurements indicate the persistence of the $N=82$ shell closure, which is consistent with the recent corresponding measurement of Yb \cite{BeckSoenke2021_Yb}.
Furthermore, the empirical shell gap at $Z=69$ continues the trend seen in the more stable $N=82$ nuclei such as $Z=63$ and 67 of having a local maximum in $\Delta_{2N}$ at the odd-$Z$ nuclei.

%The deviation in the empirical shell gap at Z=65 is most likely due to the AME value of 145Tb being a measurement of the 11/2- isomer (t_{1/2} = 31 s) rather than the ground state.

\section{Conclusion and Outlook}

%Since SMS is known to be a reliable technique, the origin of the discrepancy is unclear.

The masses of neutron deficient Tm and Er isotopes
obtained in this work identify the precise location of the proton drip-line in the Tm isotopic chain.
The transition from positive to negative proton separation energy is observed between $^{149g}$Tm and $^{150g}$Tm. 
Our updated mass of $^{149}$Er directly impacts the proton-bound nature of $^{150}$Tm since it would be the daughter nucleus of the proton emission if this decay mode were possible.
%
%, unambiguously revealing its location.
%
%allowing for an unambiguous experimental identification of its location.
%A solid understanding of the 
%%%
%%%
%The well-understood 
%composition of the mass spectra provides confidence in our identification of isotopes using the MRTOF.
%These included measurements of two previously unmeasured Tm isotopes and an update to the masses of $^{149g/m}$Er, which were obtained using the MRTOF at TRIUMF's TITAN facility.
%The location of the drip-line hinges on the strong evidence we find that the previously measured masses of $^{149g}$Er and $^{149m}$Er need to be revised, both masses lying 150$\pm$29 keV/c$^2$ higher than the current literature value, which results in $^{150}$Tm being proton-bound.
%%%
%%%
The masses of two previously unmeasured Tm isotopes were measured in this experiment. 
Additonally, an update to the previously measured mass of $^{149m}$Er and a first direct measurement of the ground state $^{149g}$Er
%The location of the drip-line hinges on the strong evidence we find that the 
determined them to be $\approx$150 keV heavier than the current literature values, thereby establishing $^{150g}$Tm as proton-bound.

%, which were obtained using the MRTOF at TRIUMF's TITAN facility.
%
The well-understood 
composition of the mass spectra provides confidence in our identification of isotopes using the MR-TOF-MS.
The mass of $^{149g}$Er was determined directly by performing hyperEMG fits on the time-of-flight spectrum obtained with TITAN's MR-TOF-MS and indirectly through the $^{149m}$Er isomer.
%Together, the measured masses strongly suggest that 
Together, our masses of $^{149g}$Tm, $^{150g}$Tm, and $^{149g}$Er establish that $^{149g}$Tm, having 20 fewer neutrons than the only stable isotope of Tm ($^{169}$Tm, $N=100$), is the first proton-unbound ground state in this isotopic chain.
This agrees with the drip-line prediction obtained from a recent Bayesian analysis of various theoretical models \cite{Tm_drip_2020}.
%This case demonstrates how a single discrepancy in a critical database can directly impact the experimental benchmarks used to test theoretical models.

%The present work suggests that %was% 150$\pm$29 keV/c$^2$

%Higher precision can be achieved by indirectly determining the ground-state mass from the isomer and the previously measured excitation energy (recent observations in the trends in the excitation energies of the (11/2-) isomers \cite{BeckSoenke2021_Yb} have added credibility to this approach). This indirect approach is how the current literature measurement was obtained.
%and the Er ground state mass is determined via the mass of the isomer, as is currently the case in the literature, but th.

%was: non-microscopic models
%As next-generation radioactive ion beam facilities begin to come online, o
Our update to the literature masses also highlights the importance of measuring and reporting not only the masses of previously unmeasured isotopes but also confirmations of literature mass values which provide input for various models and predictions.
%of both physical processes and nuclear properties.
%both physical phenomena and predictive models rely upon.
%previously measured masses
%This measurement highlights the 
%that can arise when identifying and measuring masses of 
Heavier exotic species pose challenges that arise from the abundance of isomers and isobaric molecules that must be identified and, therefore, must be treated with caution.
%Identification and measurement of heavier exotic species is a non-trivial task due to the abundance of isomers and possible isobars.
%Mass values taken from the Atomic Mass Evaluation should therefore be 
%taken with a grain of salt 
%used with caution when used to train or evaluate models.
%used with caution when used as training data for Bayesian model weighting guided by machine learning.
%likely provide misleading training data to Bayesian model weighting guided by machine learning. %Maybe change?
%For example, the deviation of $^{145}$Tb from the $S_{2p}$ trend of the isotonic chain is most likely the isomer misidentified as the ground-state, and updated measurements of such species could be of benefit.
The mass values presented here provide a new benchmark for the models used to predict the binding energies of unmeasured nuclei in the neutron-deficient region near $N=82$ and establish the $Z=69$ proton drip-line.

%2022 Nature Quote: "Proton emission from odd-odd nuclides may reveal effects of the residual proton-neutron interactions between the odd valence neutron and proton" \cite{WoodsP.J.1997NBTP}

\section{Acknowledgements}

TITAN is funded by the Natural Sciences and Engineering Research Council (NSERC) of Canada and through TRIUMF by the National Research Council (NRC) of Canada. 
%TITAN is funded by the Natural Sciences and Engineering Research Council (NSERC) of Canada and through TRIUMF by the National Research Council (NRC) of Canada. 

This work was further supported by the German Research Foundation (DFG),
Grant No.  422761894; by the German Federal
Ministry for Education and Research (BMBF), Grants
No. 05P16RGFN1 and No. 05P19RGFN1; by the
Hessian Ministry for Science and Art through the
LOEWE Center HIC for FAIR; and by the JLU and GSI
Helmholtzzentrum für Schwerionenforschung under the
JLU-GSI strategic Helmholtz partnership agreement.

%For the purpose of open access, the authors have applied a creative commons attribution (CC BY) licence to any author accepted manuscript version arising.

%We would like to thank the Targets and Ion Source group at TRIUMF. This work was supported by the Natural Sciences and Engineering Research Council (NSERC) of Canada under Grants No. SAPIN-2018-00027, No. RGPAS-2018-522453, and No. SAPPJ-2018-00028, the National Research Council (NRC) of Canada through TRIUMF, the Canada-UK Foundation, German institutions DFG (grants FR 601/3-1, SCHE 1969/2-1 and SFB 1245 and through PRISMA Cluster of Excellence), BMBF (grants 05P19RGFN1 and 05P21RGFN1), and by the JLU and GSI under the JLU-GSI strategic Helmholtz partnership agreement.

%N. M. acknowledges the support by the FAIR Phase-0 project and by Bulgarian NSF Contract No. KP-06-N48/1. E. D. acknowledges the support by the Canada-UK Foundation.

%You should: Provide one (or more) succinct conclusions that can be drawn from your data.

%For best practice in your field, spend time looking through articles in your target journal.

%Put time into abstract and conclusion!

%\section*{References}

%\begin{figure}
%\centering
%\includegraphics[width=0.85\linewidth]{images/Jan2_2024_ZhangModel.png}
%\caption[]{The measured Q-value can be used in a proton emission model such as Zhang 2018 to constrain the proton emission half-life. The expected angular momentum change of $\Delta$J$ = 5$ results in half-lives which are orders of magnitude longer than  $\Delta$J$ = 2$.}
%\label{fig:Zhang2018_half-life}
%\end{figure}
% Referenced in line 1125 %

\begin{table*}[!ht]
\centering
\caption{Data from Table \ref{table:MassSummary} without correction for unresolved isomers.
Nuclides with a {\bf *} are known or expected to have a long-lived isomer that could not be resolved in this experiment. 
No correction is made here and no additional uncertainty added to account for the admixture.
}
\small % Adjust the font size to be smaller
\begin{tabular*}{\textwidth}{@{\extracolsep{\fill}} c c c c c c}

\hline \hline \\ [-1em]
Species & Calibrant & Ion Mass Ratio & ME$_{\textrm{Measured}}$ (keV) & ME$_{\textrm{AME2020}}$ (keV) \\
[-1em]\\
\hline \\ [-1em]
%First attempt at mass ratio was 1.000048202
$^{157g+m}$Lu & $^{157}$Yb & {1.00004820(16)} & -46373 (25) & -46440 (12) \\ % mass form fit: -46373.34 keV %20.9(20) isomer
$^{157}$Tm & $^{157}$Yb & 0.99996379(18) & -58714 (28) & -58709 (28) \\ 
$^{156}$Tm & $^{156}$Yb & 0.99997554(15) & -56818 (24) & -56834 (14) \\ 
$^{155g+m}$Tm {\bf *} & $^{155}$Er & 1.00003889(15) & -56596 (23) & -56626 (10) \\ 
$^{155}$Yb & $^{155}$Er & 1.00008157(15) & -50437 (23) & -50503 (17) \\ 
$^{154g+m}$Tm {\bf *} & $^{154}$Dy & 1.00011179(18) & -54366 (26) & -54427 (14) \\ 
$^{154}$Er & $^{154}$Dy & 1.00005458(16) & -62569 (24) & -62605 (5) \\ 
$^{153g+m}$Tm {\bf *} & $^{153}$Dy & 1.00010676(36) & -53935 (52) & -53973 (12) \\ % recently ME was changed from -53955 (55) to -53957 (54) % was -53973 {\bf (34)} giving a difference of 0
$^{153}$Er & $^{153}$Dy & 1.00006064(34) & -60505 (48) & -60467 (9) \\ 
$^{152g+m}$Tm {\bf *} & $^{152}$Dy & 1.00013096(23) & -51585 (33) & -51720 (50) \\ 
$^{152}$Er & $^{152}$Dy & 1.00006789(18) & -60510 (26) & -60500 (9) \\ 
$^{151g+m}$Tm {\bf *} & $^{151}$Dy & 1.00012821(17) & -50728 (24) & -50772 (19) \\ 
$^{151}$Er & $^{151}$Dy & 1.00007472(15) & -58248 (22) & -58266 (16) \\ 
$^{150}$Tm {\bf *} & $^{150}$Dy & 1.00016386(23) & -46427 (33) & -46490 (200\#) \\ %64
$^{150}$Er & $^{150}$Dy & 1.00008237(16) & -57807 (23) & -57831 (17) \\ 
$^{149}$Tm {\bf *} & $^{149}$Dy & 1.00017251(24) & -43762 (34) & -43940 (200\#) \\ %179
$^{149g}$Er & $^{149}$Dy & 1.00010171(33) & -53584 (47) & -53742 (28) \\ 
$^{149m}$Er & $^{149}$Dy & 1.00010695(17) & -52857 (25) & -53000 (28) \\ 
$^{150}$Tm (Exp 1) {\bf *} & $^{150}$Dy & 1.00016407(45) & -46397 (63) & -46490 (200\#) \\ %93
\hline \hline
\end{tabular*}

\label{table:MassSummary_no_isomer_correction}
\end{table*}

\bibliography{references_Aug_2025.bib}

\begin{thebibliography}{10}
\expandafter\ifx\csname url\endcsname\relax
  \def\url#1{\texttt{#1}}\fi
\expandafter\ifx\csname urlprefix\endcsname\relax\def\urlprefix{URL }\fi
\expandafter\ifx\csname href\endcsname\relax
  \def\href#1#2{#2} \def\path#1{#1}\fi

\bibitem{WoodsP.J.1997NBTP}
P.~J. Woods, C.~N. Davids, {Nuclei Beyond The Proton Drip-Line}, {Annual Review of Nuclear and Particle Science} 47~(1) (1997) 541--590.

\bibitem{PfutznerM.2012Rdal}
M.~Pf\"utzner, M.~Karny, L.~V. Grigorenko, K.~Riisager, Radioactive decays at limits of nuclear stability, Rev. Mod. Phys. 84~(2) (2012) 567--619.

\bibitem{Delion2006PRL}
D.~S. Delion, R.~J. Liotta, R.~Wyss, \href{https://link.aps.org/doi/10.1103/PhysRevLett.96.072501}{{Systematics of Proton Emission}}, Phys. Rev. Lett. 96 (2006) 072501.
\newblock \href {https://doi.org/10.1103/PhysRevLett.96.072501} {\path{doi:10.1103/PhysRevLett.96.072501}}.
\newline\urlprefix\url{https://link.aps.org/doi/10.1103/PhysRevLett.96.072501}

\bibitem{Zhang2018}
Z.-X. Zhang, J.-M. Dong, \href{https://doi.org/10.1088/1674-1137/42/1/014104}{A {F}ormula for {H}alf-life of {P}roton {R}adioactivity}, Chin. Phys. C 42~(1) (2018) 014104.
\newblock \href {https://doi.org/10.1088/1674-1137/42/1/014104} {\path{doi:10.1088/1674-1137/42/1/014104}}.
\newline\urlprefix\url{https://doi.org/10.1088/1674-1137/42/1/014104}

\bibitem{SORLIN2008602}
O.~Sorlin, M.-G. Porquet, \href{https://www.sciencedirect.com/science/article/pii/S0146641008000380}{Nuclear magic numbers: New features far from stability}, Progress in Particle and Nuclear Physics 61~(2) (2008) 602--673.
\newblock \href {https://doi.org/https://doi.org/10.1016/j.ppnp.2008.05.001} {\path{doi:https://doi.org/10.1016/j.ppnp.2008.05.001}}.
\newline\urlprefix\url{https://www.sciencedirect.com/science/article/pii/S0146641008000380}

\bibitem{Kanungo_2013}
R.~Kanungo, \href{https://dx.doi.org/10.1088/0031-8949/2013/T152/014002}{A new view of nuclear shells}, Physica Scripta 2013~(T152) (2013) 014002.
\newblock \href {https://doi.org/10.1088/0031-8949/2013/T152/014002} {\path{doi:10.1088/0031-8949/2013/T152/014002}}.
\newline\urlprefix\url{https://dx.doi.org/10.1088/0031-8949/2013/T152/014002}

\bibitem{dworschak2008restoration_N82}
M.~Dworschak, G.~Audi, K.~Blaum, P.~Delahaye, S.~George, U.~Hager, F.~Herfurth, A.~Herlert, A.~Kellerbauer, H.-J. Kluge, et~al., {Restoration of the N= 82 Shell Gap from Direct Mass Measurements of Sn 132, 134}, Physical review letters 100~(7) (2008) 072501.

\bibitem{breitenfeldt2010approach_N82}
M.~Breitenfeldt, C.~Borgmann, G.~Audi, S.~Baruah, D.~Beck, K.~Blaum, C.~B{\"o}hm, R.~B. Cakirli, R.~Casten, P.~Delahaye, et~al., Approaching the n= 82 shell closure with mass measurements of ag and cd isotopes, Physical Review C—Nuclear Physics 81~(3) (2010) 034313.

\bibitem{Atanasov2015_N82}
D.~Atanasov, P.~Ascher, K.~Blaum, R.~B. Cakirli, T.~E. Cocolios, S.~George, S.~Goriely, F.~Herfurth, H.-T. Janka, O.~Just, M.~Kowalska, S.~Kreim, D.~Kisler, Y.~A. Litvinov, D.~Lunney, V.~Manea, D.~Neidherr, M.~Rosenbusch, L.~Schweikhard, A.~Welker, F.~Wienholtz, R.~N. Wolf, K.~Zuber, \href{https://link.aps.org/doi/10.1103/PhysRevLett.115.232501}{Precision mass measurements of $^{129--131}\mathrm{Cd}$ and their impact on stellar nucleosynthesis via the rapid neutron capture process}, Phys. Rev. Lett. 115 (2015) 232501.
\newblock \href {https://doi.org/10.1103/PhysRevLett.115.232501} {\path{doi:10.1103/PhysRevLett.115.232501}}.
\newline\urlprefix\url{https://link.aps.org/doi/10.1103/PhysRevLett.115.232501}

\bibitem{knoebel2016_N82}
R.~Kn{\"o}bel, M.~Diwisch, F.~Bosch, D.~Boutin, L.~Chen, C.~Dimopoulou, A.~Dolinskii, B.~Franczak, B.~Franzke, H.~Geissel, et~al., {First Direct Mass Measurements of Stored Neutron-rich 129,130,131 Cd Isotopes with FRS-ESR}, Physics Letters B 754 (2016) 288--293.

\bibitem{Babcock2018}
C.~Babcock, R.~Klawitter, E.~Leistenschneider, D.~Lascar, B.~R. Barquest, A.~Finlay, M.~Foster, A.~T. Gallant, P.~Hunt, B.~Kootte, Y.~Lan, S.~F. Paul, M.~L. Phan, M.~P. Reiter, B.~Schultz, D.~Short, C.~Andreoiu, M.~Brodeur, I.~Dillmann, G.~Gwinner, A.~A. Kwiatkowski, K.~G. Leach, J.~Dilling, \href{https://link.aps.org/doi/10.1103/PhysRevC.97.024312}{{Mass Measurements of Neutron-Rich Indium Isotopes Toward the $N=82$ Shell Closure}}, Phys. Rev. C 97 (2018) 024312.
\newblock \href {https://doi.org/10.1103/PhysRevC.97.024312} {\path{doi:10.1103/PhysRevC.97.024312}}.
\newline\urlprefix\url{https://link.aps.org/doi/10.1103/PhysRevC.97.024312}

\bibitem{PhysRevLett.124.092502}
V.~Manea, J.~Karthein, D.~Atanasov, M.~Bender, K.~Blaum, T.~E. Cocolios, S.~Eliseev, A.~Herlert, J.~D. Holt, W.~J. Huang, Y.~A. Litvinov, D.~Lunney, J.~Men\'endez, M.~Mougeot, D.~Neidherr, L.~Schweikhard, A.~Schwenk, J.~Simonis, A.~Welker, F.~Wienholtz, K.~Zuber, \href{https://link.aps.org/doi/10.1103/PhysRevLett.124.092502}{First glimpse of the $n=82$ shell closure below $z=50$ from masses of neutron-rich cadmium isotopes and isomers}, Phys. Rev. Lett. 124 (2020) 092502.
\newblock \href {https://doi.org/10.1103/PhysRevLett.124.092502} {\path{doi:10.1103/PhysRevLett.124.092502}}.
\newline\urlprefix\url{https://link.aps.org/doi/10.1103/PhysRevLett.124.092502}

\bibitem{beck2000accurate}
D.~Beck, F.~Ames, G.~Audi, G.~Bollen, F.~Herfurth, H.~J. Kluge, A.~Kohl, M.~K{\"o}nig, D.~Lunney, I.~Martel, et~al., {Accurate Masses of Unstable Rare-earth Isotopes by ISOLTRAP}, The European Physical Journal A 8 (2000) 307--329.

\bibitem{2005Li24}
Y.~A. Litvinov, H.~Geissel, T.~Radon, F.~Attallah, G.~Audi, K.~Beckert, F.~Bosch, M.~Falch, B.~Franzke, M.~Hausmann, et~al., {Mass Measurement of Cooled Neutron-Deficient Bismuth Projectile Fragments with Time-Resolved Schottky Mass Spectrometry at the FRS-ESR Facility}, Nuclear Physics A 756~(1-2) (2005) 3--38.

\bibitem{BeckSoenke2021_Yb}
S.~Beck, B.~Kootte, I.~Dedes, T.~Dickel, A.~A. Kwiatkowski, E.~M. Lykiardopoulou, W.~R. Plaß, M.~P. Reiter, C.~Andreoiu, J.~Bergmann, T.~Brunner, D.~Curien, J.~Dilling, J.~Dudek, E.~Dunling, J.~Flowerdew, A.~Gaamouci, L.~Graham, G.~Gwinner, A.~Jacobs, R.~Klawitter, Y.~Lan, E.~Leistenschneider, N.~Minkov, V.~Monier, I.~Mukul, S.~F. Paul, C.~Scheidenberger, R.~I. Thompson, J.~L. Tracy, M.~Vansteenkiste, H.-L. Wang, M.~E. Wieser, C.~Will, J.~Yang, {Mass Measurements of Neutron-Deficient Yb Isotopes and Nuclear Structure at the Extreme Proton-Rich Side of the N=82 Shell}, Physical review letters 127~(11) (2021) 112501--112501.

\bibitem{WOLLNIK_MRTOF_1990}
H.~Wollnik, M.~Przewloka, \href{https://www.sciencedirect.com/science/article/pii/016811769085127N}{Time-of-flight mass spectrometers with multiply reflected ion trajectories}, International Journal of Mass Spectrometry and Ion Processes 96~(3) (1990) 267--274.
\newblock \href {https://doi.org/https://doi.org/10.1016/0168-1176(90)85127-N} {\path{doi:https://doi.org/10.1016/0168-1176(90)85127-N}}.
\newline\urlprefix\url{https://www.sciencedirect.com/science/article/pii/016811769085127N}

\bibitem{delion2021universal}
D.~Delion, A.~Dumitrescu, {Universal Proton Emission Systematics}, Physical Review C 103~(5) (2021) 054325.

\bibitem{geng2004proton}
L.~Geng, H.~Toki, J.~Meng, Proton-rich nuclei at and beyond the proton drip line in the relativistic mean field theory, Progress of theoretical physics 112~(4) (2004) 603--617.

\bibitem{Rauth2008_Penning_beyond_p_drip}
C.~Rauth, D.~Ackermann, K.~Blaum, M.~Block, A.~Chaudhuri, Z.~Di, S.~Eliseev, R.~Ferrer, D.~Habs, F.~Herfurth, F.~P. He\ss{}berger, S.~Hofmann, H.-J. Kluge, G.~Maero, A.~Mart\'{\i}n, G.~Marx, M.~Mukherjee, J.~B. Neumayr, W.~R. Pla\ss{}, S.~Rahaman, D.~Rodr\'{\i}guez, C.~Scheidenberger, L.~Schweikhard, P.~G. Thirolf, G.~Vorobjev, C.~Weber, \href{https://link.aps.org/doi/10.1103/PhysRevLett.100.012501}{{First Penning Trap Mass Measurements beyond the Proton Drip Line}}, Phys. Rev. Lett. 100 (2008) 012501.
\newblock \href {https://doi.org/10.1103/PhysRevLett.100.012501} {\path{doi:10.1103/PhysRevLett.100.012501}}.
\newline\urlprefix\url{https://link.aps.org/doi/10.1103/PhysRevLett.100.012501}

\bibitem{hofmann1982proton}
S.~Hofmann, W.~Reisdorf, G.~M{\"u}nzenberg, F.~He{\ss}berger, J.~Schneider, P.~Armbruster, Proton radioactivity of $^{151}${L}u, Zeitschrift f{\"u}r Physik A Atoms and Nuclei 305~(2) (1982) 111--123.

\bibitem{klepper1982direct}
O.~Klepper, T.~Batsch, S.~Hofmann, R.~Kirchner, W.~Kurcewicz, W.~Reisdorf, E.~Roeckl, D.~Schardt, G.~Nyman, Direct and beta-delayed proton decay of very neutron-deficient rare-earth isotopes produced in the reaction $^{58}${N}i + $^{92}${M}o, Zeitschrift f{\"u}r Physik A Atoms and Nuclei 305~(2) (1982) 125--130.

\bibitem{BLANK2008403}
B.~Blank, M.~Borge, \href{https://www.sciencedirect.com/science/article/pii/S0146641007000956}{{Nuclear structure at the proton drip line: Advances with nuclear decay studies}}, Progress in Particle and Nuclear Physics 60~(2) (2008) 403--483.
\newblock \href {https://doi.org/https://doi.org/10.1016/j.ppnp.2007.12.001} {\path{doi:https://doi.org/10.1016/j.ppnp.2007.12.001}}.
\newline\urlprefix\url{https://www.sciencedirect.com/science/article/pii/S0146641007000956}

\bibitem{ENSDF-webpage}
\href{http://www.nndc.bnl.gov/ensdf/}{{Evaluated Nuclear Structure Data Files of Brookhaven National Laboratory}} (2021).
\newline\urlprefix\url{http://www.nndc.bnl.gov/ensdf/}

\bibitem{zhang2022observation}
W.~Zhang, B.~Cederwall, {\"O}.~Aktas, X.~Liu, A.~Ertoprak, A.~Nyberg, K.~Auranen, B.~Alayed, H.~Badran, H.~Boston, et~al., Observation of the proton emitter $^{116}_{57}${L}a$_{59}$, Communications Physics 5 (2022).

\bibitem{ERLERJochen2012Tlot}
J.~Erler, N.~Birge, M.~Kortelainen, W.~Nazarewicz, E.~Olsen, A.~M. Perhac, M.~Stoitsov, The {L}imits of the {N}uclear {L}andscape, Nature (London) 486~(7404) (2012) 509--512.

\bibitem{Kortelainen2012UNEDF1}
M.~Kortelainen, J.~McDonnell, W.~Nazarewicz, P.-G. Reinhard, J.~Sarich, N.~Schunck, M.~V. Stoitsov, S.~M. Wild, \href{https://link.aps.org/doi/10.1103/PhysRevC.85.024304}{{Nuclear Energy Density Optimization: Large Deformations}}, Phys. Rev. C 85 (2012) 024304.
\newblock \href {https://doi.org/10.1103/PhysRevC.85.024304} {\path{doi:10.1103/PhysRevC.85.024304}}.
\newline\urlprefix\url{https://link.aps.org/doi/10.1103/PhysRevC.85.024304}

\bibitem{Klupfel2009_SV-min}
P.~Kl\"upfel, P.-G. Reinhard, T.~J. B\"urvenich, J.~A. Maruhn, \href{https://link.aps.org/doi/10.1103/PhysRevC.79.034310}{{Variations on a theme by Skyrme: A systematic study of adjustments of model parameters}}, Phys. Rev. C 79 (2009) 034310.
\newblock \href {https://doi.org/10.1103/PhysRevC.79.034310} {\path{doi:10.1103/PhysRevC.79.034310}}.
\newline\urlprefix\url{https://link.aps.org/doi/10.1103/PhysRevC.79.034310}

\bibitem{BSkG3_grams2023skyrme}
G.~Grams, W.~Ryssens, G.~Scamps, S.~Goriely, N.~Chamel, Skyrme-hartree-fock-bogoliubov mass models on a 3d mesh: Iii. from atomic nuclei to neutron stars, The European Physical Journal A 59~(11) (2023) 270.

\bibitem{Kejzlar2020statisticalBE}
V.~Kejzlar, L.~Neufcourt, W.~Nazarewicz, P.-G. Reinhard, \href{https://dx.doi.org/10.1088/1361-6471/ab907c}{Statistical aspects of nuclear mass models}, Journal of Physics G: Nuclear and Particle Physics 47~(9) (2020) 094001.
\newblock \href {https://doi.org/10.1088/1361-6471/ab907c} {\path{doi:10.1088/1361-6471/ab907c}}.
\newline\urlprefix\url{https://dx.doi.org/10.1088/1361-6471/ab907c}

\bibitem{Tm_drip_2020}
L.~Neufcourt, Y.~Cao, S.~Giuliani, W.~Nazarewicz, E.~Olsen, O.~B. Tarasov, \href{https://link.aps.org/doi/10.1103/PhysRevC.101.014319}{{Beyond the Proton Drip Line: Bayesian Analysis of Proton-Emitting Nuclei}}, Phys. Rev. C 101 (2020) 014319.
\newblock \href {https://doi.org/10.1103/PhysRevC.101.014319} {\path{doi:10.1103/PhysRevC.101.014319}}.
\newline\urlprefix\url{https://link.aps.org/doi/10.1103/PhysRevC.101.014319}

\bibitem{Ball_2016}
G.~C. Ball, G.~Hackman, R.~Krücken, \href{https://dx.doi.org/10.1088/0031-8949/91/9/093002}{The triumf-isac facility: two decades of discovery with rare isotope beams}, Physica Scripta 91~(9) (2016) 093002.
\newblock \href {https://doi.org/10.1088/0031-8949/91/9/093002} {\path{doi:10.1088/0031-8949/91/9/093002}}.
\newline\urlprefix\url{https://dx.doi.org/10.1088/0031-8949/91/9/093002}

\bibitem{BRICAULT200249}
P.~Bricault, R.~Baartman, M.~Dombsky, A.~Hurst, C.~Mark, G.~Stanford, P.~Schmor, \href{https://www.sciencedirect.com/science/article/pii/S0375947401015469}{{TRIUMF-ISAC Target Station and Mass Separator Commissioning}}, Nuclear Physics A 701~(1) (2002) 49--53, 5th International Conference on Radioactive Nuclear Beams.
\newblock \href {https://doi.org/https://doi.org/10.1016/S0375-9474(01)01546-9} {\path{doi:https://doi.org/10.1016/S0375-9474(01)01546-9}}.
\newline\urlprefix\url{https://www.sciencedirect.com/science/article/pii/S0375947401015469}

\bibitem{BRUNNER2012_RFQ}
T.~Brunner, M.~Smith, M.~Brodeur, S.~Ettenauer, A.~Gallant, V.~Simon, A.~Chaudhuri, A.~Lapierre, E.~Mané, R.~Ringle, M.~Simon, J.~Vaz, P.~Delheij, M.~Good, M.~Pearson, J.~Dilling, \href{https://www.sciencedirect.com/science/article/pii/S0168900212001398}{{TITAN's digital RFQ ion beam cooler and buncher, operation and performance}}, Nuclear Instruments and Methods in Physics Research Section A: Accelerators, Spectrometers, Detectors and Associated Equipment 676 (2012) 32--43.
\newblock \href {https://doi.org/https://doi.org/10.1016/j.nima.2012.02.004} {\path{doi:https://doi.org/10.1016/j.nima.2012.02.004}}.
\newline\urlprefix\url{https://www.sciencedirect.com/science/article/pii/S0168900212001398}

\bibitem{jesch2017mrTOF}
C.~Jesch, T.~Dickel, W.~R. Pla{\ss}, D.~Short, S.~A.~S. Andres, J.~Dilling, H.~Geissel, F.~Greiner, J.~Lang, K.~G. Leach, W.~Lippert, C.~Scheidenberger, M.~I. Yavor, {The MR-TOF-MS isobar separator for the TITAN facility at TRIUMF}, in: M.~Wada, P.~Schury, Y.~Ichikawa (Eds.), TCP 2014, Springer International Publishing, Cham, 2017, pp. 175--184.

\bibitem{dickel2019recent}
T.~Dickel, S.~A. San~Andr{\'e}s, S.~Beck, J.~Bergmann, J.~Dilling, F.~Greiner, C.~Hornung, A.~Jacobs, G.~Kripko-Koncz, A.~Kwiatkowski, et~al., \href{https://link.springer.com/article/10.1007%2Fs10751-019-1610-y}{{Recent Upgrades of the Multiple-Reflection Time-of-Flight Mass Spectrometer at TITAN, TRIUMF}}, Hyperfine Interactions 240~(1) (2019) 1--9.
\newline\urlprefix\url{https://link.springer.com/article/10.1007%2Fs10751-019-1610-y}

\bibitem{REITER2021MRTOF_commissioning}
M.~Reiter, S.~A.~S. Andrés, J.~Bergmann, T.~Dickel, J.~Dilling, A.~Jacobs, A.~Kwiatkowski, W.~Plaß, C.~Scheidenberger, D.~Short, C.~Will, C.~Babcock, E.~Dunling, A.~Finlay, C.~Hornung, C.~Jesch, R.~Klawitter, B.~Kootte, D.~Lascar, E.~Leistenschneider, T.~Murböck, S.~Paul, M.~Yavor, \href{https://www.sciencedirect.com/science/article/pii/S0168900221008081}{{Commissioning and Performance of TITAN’s Multiple-Reflection Time-of-Flight Mass-Spectrometer and Isobar Separator}}, Nuclear Instruments and Methods in Physics Research Section A: Accelerators, Spectrometers, Detectors and Associated Equipment 1018 (2021) 165823.
\newblock \href {https://doi.org/https://doi.org/10.1016/j.nima.2021.165823} {\path{doi:https://doi.org/10.1016/j.nima.2021.165823}}.
\newline\urlprefix\url{https://www.sciencedirect.com/science/article/pii/S0168900221008081}

\bibitem{DICKEL2015172_MRS}
T.~Dickel, W.~Plaß, A.~Becker, U.~Czok, H.~Geissel, E.~Haettner, C.~Jesch, W.~Kinsel, M.~Petrick, C.~Scheidenberger, A.~Simon, M.~Yavor, \href{https://www.sciencedirect.com/science/article/pii/S0168900214015629}{A high-performance multiple-reflection time-of-flight mass spectrometer and isobar separator for the research with exotic nuclei}, Nuclear Instruments and Methods in Physics Research Section A: Accelerators, Spectrometers, Detectors and Associated Equipment 777 (2015) 172--188.
\newblock \href {https://doi.org/https://doi.org/10.1016/j.nima.2014.12.094} {\path{doi:https://doi.org/10.1016/j.nima.2014.12.094}}.
\newline\urlprefix\url{https://www.sciencedirect.com/science/article/pii/S0168900214015629}

\bibitem{DICKEL20171_retrapping}
T.~Dickel, M.~I. Yavor, J.~Lang, W.~R. Plaß, W.~Lippert, H.~Geissel, C.~Scheidenberger, \href{https://www.sciencedirect.com/science/article/pii/S1387380616302664}{Dynamical time focus shift in multiple-reflection time-of-flight mass spectrometers}, International Journal of Mass Spectrometry 412 (2017) 1--7.
\newblock \href {https://doi.org/https://doi.org/10.1016/j.ijms.2016.11.005} {\path{doi:https://doi.org/10.1016/j.ijms.2016.11.005}}.
\newline\urlprefix\url{https://www.sciencedirect.com/science/article/pii/S1387380616302664}

\bibitem{dickel2017isobar}
T.~Dickel, W.~R. Pla{\ss}, W.~Lippert, J.~Lang, M.~I. Yavor, H.~Geissel, C.~Scheidenberger, {Isobar Separation in a Multiple-Reflection Time-of-Flight Mass Spectrometer by Mass-Selective Re-Trapping}, Journal of The American Society for Mass Spectrometry 28~(6) (2017) 1079--1090.

\bibitem{Lykiard_pEmitters_2023}
E.~M. Lykiardopoulou, G.~Audi, T.~Dickel, W.~J. Huang, D.~Lunney, W.~R. Pla\ss{}, M.~P. Reiter, J.~Dilling, A.~A. Kwiatkowski, \href{https://link.aps.org/doi/10.1103/PhysRevC.107.024311}{Exploring the limits of existence of proton-rich nuclei in the $z=70--82$ region}, Phys. Rev. C 107 (2023) 024311.
\newblock \href {https://doi.org/10.1103/PhysRevC.107.024311} {\path{doi:10.1103/PhysRevC.107.024311}}.
\newline\urlprefix\url{https://link.aps.org/doi/10.1103/PhysRevC.107.024311}

\bibitem{Ayet2019_systematics}
S.~Ayet San~Andr\'es, C.~Hornung, J.~Ebert, W.~R. Pla\ss{}, T.~Dickel, H.~Geissel, C.~Scheidenberger, J.~Bergmann, F.~Greiner, E.~Haettner, C.~Jesch, W.~Lippert, I.~Mardor, I.~Miskun, Z.~Patyk, S.~Pietri, A.~Pihktelev, S.~Purushothaman, M.~P. Reiter, A.-K. Rink, H.~Weick, M.~I. Yavor, S.~Bagchi, V.~Charviakova, P.~Constantin, M.~Diwisch, A.~Finlay, S.~Kaur, R.~Kn\"obel, J.~Lang, B.~Mei, I.~D. Moore, J.-H. Otto, I.~Pohjalainen, A.~Prochazka, C.~Rappold, M.~Takechi, Y.~K. Tanaka, J.~S. Winfield, X.~Xu, \href{https://link.aps.org/doi/10.1103/PhysRevC.99.064313}{High-resolution, accurate multiple-reflection time-of-flight mass spectrometry for short-lived, exotic nuclei of a few events in their ground and low-lying isomeric states}, Phys. Rev. C 99 (2019) 064313.
\newblock \href {https://doi.org/10.1103/PhysRevC.99.064313} {\path{doi:10.1103/PhysRevC.99.064313}}.
\newline\urlprefix\url{https://link.aps.org/doi/10.1103/PhysRevC.99.064313}

\bibitem{EMG2017}
S.~Purushothaman, S.~{Ayet San Andrés}, J.~Bergmann, T.~Dickel, J.~Ebert, H.~Geissel, C.~Hornung, W.~Plaß, C.~Rappold, C.~Scheidenberger, Y.~Tanaka, M.~Yavor, \href{http://www.sciencedirect.com/science/article/pii/S1387380616302913}{{Hyper-EMG: A New Probability Distribution Function Composed of Exponentially Modified Gaussian Distributions to Analyze Asymmetric Peak Shapes in High-Resolution Time-of-Flight Mass Spectrometry}}, International Journal of Mass Spectrometry 421 (2017) 245 -- 254.
\newblock \href {https://doi.org/https://doi.org/10.1016/j.ijms.2017.07.014} {\path{doi:https://doi.org/10.1016/j.ijms.2017.07.014}}.
\newline\urlprefix\url{http://www.sciencedirect.com/science/article/pii/S1387380616302913}

\bibitem{EMGfit}
S.~F. Paul, \href{https://pypi.org/project/emgfit/}{{emgfit 0.5.0}} (2025).
\newline\urlprefix\url{https://pypi.org/project/emgfit/}

\bibitem{Reiter2018}
M.~P. Reiter, S.~Ayet San~Andr\'es, E.~Dunling, B.~Kootte, E.~Leistenschneider, C.~Andreoiu, C.~Babcock, B.~R. Barquest, J.~Bollig, T.~Brunner, I.~Dillmann, A.~Finlay, G.~Gwinner, L.~Graham, J.~D. Holt, C.~Hornung, C.~Jesch, R.~Klawitter, Y.~Lan, D.~Lascar, J.~E. McKay, S.~F. Paul, R.~Steinbr\"ugge, R.~Thompson, J.~L. Tracy, M.~E. Wieser, C.~Will, T.~Dickel, W.~R. Pla\ss{}, C.~Scheidenberger, A.~A. Kwiatkowski, J.~Dilling, \href{https://link.aps.org/doi/10.1103/PhysRevC.98.024310}{{Quenching of the $N=32$ Neutron Shell Closure Studied via Precision Mass Measurements of Neutron-Rich Vanadium Isotopes}}, Phys. Rev. C 98 (2018) 024310.
\newblock \href {https://doi.org/10.1103/PhysRevC.98.024310} {\path{doi:10.1103/PhysRevC.98.024310}}.
\newline\urlprefix\url{https://link.aps.org/doi/10.1103/PhysRevC.98.024310}

\bibitem{Kortelahti1989_153Tm_isomer}
M.~O. Kortelahti, K.~S. Toth, K.~S. Vierinen, J.~M. Nitschke, P.~A. Wilmarth, R.~B. Firestone, R.~M. Chasteler, A.~A. Shihab-Eldin, \href{https://link.aps.org/doi/10.1103/PhysRevC.39.636}{Decay properties of $^{153}\mathrm{Yb}$ and $^{153}\mathrm{Tm}$; excitation energies of the ${s}_{1/2}$ and ${h}_{11/2}$ proton states in $^{153}\mathrm{Tm}$}, Phys. Rev. C 39 (1989) 636--644.
\newblock \href {https://doi.org/10.1103/PhysRevC.39.636} {\path{doi:10.1103/PhysRevC.39.636}}.
\newline\urlprefix\url{https://link.aps.org/doi/10.1103/PhysRevC.39.636}

\bibitem{potempa1990h}
A.~Potempa, V.~Afanas'ev, Y.~Vavryshchuk, $h_{11/2}$ and $s_{1/2}$ isomeric states in $^{155}$tm., Izvestiya Akademii Nauk SSSR, Seriya Fizicheskaya;(USSR) 54~(5) (1990).

\bibitem{SINGH2009_ENSDF}
B.~Singh, \href{https://www.sciencedirect.com/science/article/pii/S0090375208001300}{Nuclear data sheets for a = 151}, Nuclear Data Sheets 110~(1) (2009) 1--264.
\newblock \href {https://doi.org/https://doi.org/10.1016/j.nds.2008.11.035} {\path{doi:https://doi.org/10.1016/j.nds.2008.11.035}}.
\newline\urlprefix\url{https://www.sciencedirect.com/science/article/pii/S0090375208001300}

\bibitem{martinNDC_2013}
{Martin, M.J.}, {Adopted levels, gammas for 152Sm}, {Nucl. Data Sheets} 114 (2013) 1497.

\bibitem{ENSDF_154Gd_reich2009}
C.~Reich, \href{https://www.sciencedirect.com/science/article/pii/S0090375209000805}{Nuclear data sheets for a = 154}, Nuclear Data Sheets 110~(10) (2009) 2257--2532.
\newblock \href {https://doi.org/https://doi.org/10.1016/j.nds.2009.09.001} {\path{doi:https://doi.org/10.1016/j.nds.2009.09.001}}.
\newline\urlprefix\url{https://www.sciencedirect.com/science/article/pii/S0090375209000805}

\bibitem{Kondev2021_nubase2020}
F.~Kondev, M.~Wang, W.~Huang, S.~Naimi, G.~Audi, \href{https://dx.doi.org/10.1088/1674-1137/abddae}{{The NUBASE2020 evaluation of nuclear physics properties *}}, Chinese Physics C 45~(3) (2021) 030001.
\newblock \href {https://doi.org/10.1088/1674-1137/abddae} {\path{doi:10.1088/1674-1137/abddae}}.
\newline\urlprefix\url{https://dx.doi.org/10.1088/1674-1137/abddae}

\bibitem{149TmIsomer_broda1987level}
R.~Broda, P.~Daly, J.~McNeill, R.~Janssens, D.~Radford, {Level structure of 68 149 Er 81 and high-spin isomerism in proton-rich N= 81, 82, 83 nuclei}, Zeitschrift f{\"u}r Physik A Atomic Nuclei 327 (1987) 403--408.

\bibitem{150TmIsomer_PhysRevC.37.2694}
J.~M. Nitschke, P.~A. Wilmarth, J.~Gilat, K.~S. Toth, F.~T. Avignone, \href{https://link.aps.org/doi/10.1103/PhysRevC.37.2694}{{Delayed proton emission of N=81 odd-odd precursors: $^{148}\mathrm{Ho}$, $^{150}\mathrm{Tm}$ , and $^{152}\mathrm{Lu}$}}, Phys. Rev. C 37 (1988) 2694--2703.
\newblock \href {https://doi.org/10.1103/PhysRevC.37.2694} {\path{doi:10.1103/PhysRevC.37.2694}}.
\newline\urlprefix\url{https://link.aps.org/doi/10.1103/PhysRevC.37.2694}

\bibitem{potempa1992investigation}
A.~Potempa, K.~Y. Gromov, J.~Wawryszczuk, V.~Kalinnikov, {Investigation of $\alpha$-Decay of Spherical Nucleus Isomers with $Z > 64$}, Bulletin - Russian Academy of Sciences: Physics 56 (1992) 666--666.

\bibitem{AME2020-I}
W.~Huang, M.~Wang, F.~G. Kondev, G.~Audi, S.~Naimi, {The AME 2020 Atomic Mass Evaluation (I). Evaluation of Input Data; and Adjustment Procedures}, Chinese Physics C: High Energy Physics and Nuclear Physics 45~(3) (2021) -- , 030002, 10.1088/1674-1137/abddb0.

\bibitem{rauth2007direct}
C.~Rauth, D.~Ackermann, G.~Audi, M.~Block, A.~Chaudhuri, S.~Eliseev, F.~Herfurth, F.~Hessberger, S.~Hofmann, H.-J. Kluge, et~al., {Direct mass measurements around A= 146 at SHIPTRAP}, The European Physical Journal Special Topics 150 (2007) 329--335.

\bibitem{Block2007}
M.~Block, D.~Ackermann, K.~Blaum, A.~Chaudhuri, Z.~Di, S.~Eliseev, R.~Ferrer, D.~Habs, F.~Herfurth, F.~P. Heßberger, S.~Hofmann, H.~Kluge, G.~Maero, A.~Martín, G.~Marx, M.~Mazzocco, M.~Mukherjee, J.~B. Neumayr, W.~R. Plaß, W.~Quint, S.~Rahaman, C.~Rauth, D.~Rodríguez, C.~Scheidenberger, L.~Schweikhard, P.~G. Thirolf, G.~Vorobjev, C.~Weber, \href{https://doi.org/10.1063/1.2746619}{{Mass measurements of exotic nuclides at SHIPTRAP}}, AIP Conference Proceedings 912~(1) (2007) 423--430.
\newblock \href {http://arxiv.org/abs/https://pubs.aip.org/aip/acp/article-pdf/912/1/423/11695800/423\_1\_online.pdf} {\path{arXiv:https://pubs.aip.org/aip/acp/article-pdf/912/1/423/11695800/423\_1\_online.pdf}}, \href {https://doi.org/10.1063/1.2746619} {\path{doi:10.1063/1.2746619}}.
\newline\urlprefix\url{https://doi.org/10.1063/1.2746619}

\bibitem{Toth1985}
K.~S. Toth, Y.~A. Ellis-Akovali, F.~T. Avignone, R.~S. Moore, D.~M. Moltz, J.~M. Nitschke, P.~A. Wilmarth, P.~K. Lemmertz, D.~C. Sousa, A.~L. Goodman, \href{https://link.aps.org/doi/10.1103/PhysRevC.32.342}{{Single-Particle States in $^{149}$Er and $^{149}$Ho, and the Effect of the Z=64 Closure}}, Phys. Rev. C 32 (1985) 342--345.
\newblock \href {https://doi.org/10.1103/PhysRevC.32.342} {\path{doi:10.1103/PhysRevC.32.342}}.
\newline\urlprefix\url{https://link.aps.org/doi/10.1103/PhysRevC.32.342}

\bibitem{Firestone1989}
R.~B. Firestone, J.~M. Nitschke, P.~A. Wilmarth, K.~Vierinen, J.~Gilat, K.~S. Toth, Y.~A. Akovali, \href{https://link.aps.org/doi/10.1103/PhysRevC.39.219}{{Decay of $^{149}\mathrm{Er}^{\mathrm{g}+\mathrm{m}}$ by Positron and Delayed Proton Emission and by Electron Capture}}, Phys. Rev. C 39 (1989) 219--232.
\newblock \href {https://doi.org/10.1103/PhysRevC.39.219} {\path{doi:10.1103/PhysRevC.39.219}}.
\newline\urlprefix\url{https://link.aps.org/doi/10.1103/PhysRevC.39.219}

\bibitem{AME2020-II}
M.~Wang, W.~Huang, F.~Kondev, G.~Audi, S.~Naimi, \href{https://doi.org/10.1088/1674-1137/abddaf}{{The {AME} 2020 Atomic Mass Evaluation ({II}). Tables, Graphs and References}}, Chinese Physics C 45~(3) (2021) 030003.
\newblock \href {https://doi.org/10.1088/1674-1137/abddaf} {\path{doi:10.1088/1674-1137/abddaf}}.
\newline\urlprefix\url{https://doi.org/10.1088/1674-1137/abddaf}

\bibitem{Kortelainen2010UNEDF0}
M.~Kortelainen, T.~Lesinski, J.~Mor{\'e}, W.~Nazarewicz, J.~Sarich, N.~Schunck, M.~Stoitsov, S.~Wild, Nuclear energy density optimization, Physical Review C—Nuclear Physics 82~(2) (2010) 024313.

\bibitem{goriely2010HFB21}
S.~Goriely, N.~Chamel, J.~Pearson, Further explorations of skyrme-hartree-fock-bogoliubov mass formulas. xii. stiffness and stability of neutron-star matter, Physical Review C—Nuclear Physics 82~(3) (2010) 035804.

\bibitem{duflo-zuker1995microscopic}
J.~Duflo, A.~Zuker, Microscopic mass formulas, Physical Review C 52~(1) (1995) R23.

\bibitem{moller1988FRDM}
P.~M{\"o}ller, W.~Myers, W.~Swiatecki, J.~Treiner, Nuclear mass formula with a finite-range droplet model and a folded-yukawa single-particle potential, Atomic data and nuclear data tables 39~(2) (1988) 225--233.

\bibitem{dobaczewski1984_SkP}
J.~Dobaczewski, H.~Flocard, J.~Treiner, Hartree-fock-bogolyubov description of nuclei near the neutron-drip line, Nuclear Physics A 422~(1) (1984) 103--139.

\bibitem{pearson1996_ETFSI-Q}
J.~Pearson, R.~Nayak, S.~Goriely, Nuclear mass formula with bogolyubov-enhanced shell-quenching: application to r-process, Physics Letters B 387~(3) (1996) 455--459.

\bibitem{Izzo2021_In}
C.~Izzo, J.~Bergmann, K.~A. Dietrich, E.~Dunling, D.~Fusco, A.~Jacobs, B.~Kootte, G.~Kripk\'o-Koncz, Y.~Lan, E.~Leistenschneider, E.~M. Lykiardopoulou, I.~Mukul, S.~F. Paul, M.~P. Reiter, J.~L. Tracy, C.~Andreoiu, T.~Brunner, T.~Dickel, J.~Dilling, I.~Dillmann, G.~Gwinner, D.~Lascar, K.~G. Leach, W.~R. Pla\ss{}, C.~Scheidenberger, M.~E. Wieser, A.~A. Kwiatkowski, \href{https://link.aps.org/doi/10.1103/PhysRevC.103.025811}{Mass measurements of neutron-rich indium isotopes for $r$-process studies}, Phys. Rev. C 103 (2021) 025811.
\newblock \href {https://doi.org/10.1103/PhysRevC.103.025811} {\path{doi:10.1103/PhysRevC.103.025811}}.
\newline\urlprefix\url{https://link.aps.org/doi/10.1103/PhysRevC.103.025811}

\end{thebibliography}
\bibliographystyle{elsarticle-num}

\end{document}